\documentclass[11pt]{article}
\usepackage{amssymb}

\usepackage[totalwidth=460truept,totalheight=600truept]{geometry}
\usepackage{latexsym,amssymb,amsmath,graphicx,accents,eucal,slashed,subfigure}
\usepackage{dsfont}
\usepackage[T1]{fontenc}
\usepackage{hyperref}

\def\theequation{\arabic{section}.\arabic{equation}}

\renewcommand{\theequation}{\thesection.\arabic{equation}}
\global\arraycolsep=1truept

\numberwithin{equation}{section}
\renewcommand{\theequation}{\arabic{section}.\arabic{equation}}
\newtheorem{theorem}{Theorem}

\begin{document}

\bigskip \hfill IFUP-TH 2012/09

\vskip 1.4truecm

\begin{center}
{\huge \textbf{A Master Functional}}

\vskip .5truecm

{\huge \textbf{For Quantum Field Theory}}

\vskip 1truecm

\textsl{Damiano Anselmi} \vskip .2truecm

\textit{Dipartimento di Fisica ``Enrico Fermi'', Universit\`{a} di Pisa, }

\textit{Largo B. Pontecorvo 3, I-56127 Pisa, Italy,}

\vskip .2truecm

damiano.anselmi@df.unipi.it

\vskip 1.5truecm

\textbf{Abstract}
\end{center}

\medskip

{\small We study a new generating functional of one-particle irreducible
diagrams in quantum field theory, called master functional, which is
invariant under the most general perturbative changes of field variables.
The usual} {\small functional }$\Gamma ${\small \ does not behave as a
scalar under the transformation law inherited from its very definition as
the Lagendre transform of }$W=\ln Z${\small , although it does behave as a
scalar under an unusual transformation law. The master functional, on the
other hand, is the Legendre transform of an improved functional }$W${\small %
\ with respect to the sources coupled to both elementary and composite
fields. The inclusion of certain improvement terms in }$W${\small \ and }$Z$%
{\small \ is necessary to make the new Legendre transform well defined. The
master functional behaves as a scalar under the transformation law inherited
from its very definition. Moreover, it admits a proper formulation, obtained
extending the set of integrated fields to so-called proper fields, which
allows us to work without passing through }$Z${\small , }$W${\small \ or }$%
\Gamma ${\small . In the proper formulation the classical action coincides
with the classical limit of the master functional, and correlation functions
and renormalization are calculated applying the usual diagrammatic rules to
the proper fields. Finally, the most general change of field variables,
including the map relating bare and renormalized fields, is a linear
redefinition of the proper fields.}

\vskip 1truecm

\vfill\eject

\section{Introduction}

\setcounter{equation}{0}

Renormalization, as it is usually formulated, is not a change of variables
in the functional integral, combined with parameter redefinitions, but a
simple replacement of variables and parameters inside the action. More
precisely, the action is correctly transformed according to the field
redefinition, but the term $\int J\varphi $, which identifies the
``elementary field'' used to write Feynman rules and calculate diagrams, is
not transformed, rather just replaced with $\int J^{\prime }\varphi ^{\prime
}$, the analogous term for the new variables. In simple power-counting
renormalizable theories, such as ordinary Yang-Mills theory, where the
renormalization of fields and sources is multiplicative, it is
straightforward to turn replacements into true changes of field variables.
Instead, in theories such as Yang-Mills theory with an unusual action, or
with composite fields turned on, as well as effective field theories and
gravity, the relation between bare and renormalized fields can be non-linear 
\cite{thooftveltman}. In those cases replacements are convenient shortcuts
that allow us to avoid certain lengthy manipulations. However, they are not
completely satisfactory, since the do not really allow us to write precise
identities relating generating functionals before and after the changes of
variables.

A perturbative field redefinition is a field redefinition that can be
expressed as the identity map plus a perturbative series of local monomials
of the fields and their derivatives. In ref. \cite{fieldcov} we studied how
a general perturbative change of integration variables in the functional
integral reflects on the generating functionals $Z$ and $W=\ln Z$. Due to
the intimate relation between composite fields $\mathcal{O}^{I}(\varphi )$
and changes of field variables, it is convenient to include sources $L_{I}$
coupled to the $\mathcal{O}^{I}(\varphi )$s, besides the sources $J$ coupled
to the elementary fields $\varphi $. In a particularly convenient approach,
called \textit{linear approach}, all perturbative changes of field
variables, including the \textit{BR map}, which is the map relating bare and
renormalized fields, are expressed as linear source redefinitions of the
form 
\begin{equation}
L_{I}=L_{J}^{\prime }z_{I}^{J}+b_{I}J^{\prime },\qquad J=J^{\prime },
\label{bibip}
\end{equation}
where $z_{I}^{J}$ and $b_{I}$ are constants. The $Z$- and $W$- functionals
behave as scalars, 
\begin{equation}
Z^{\prime }(J^{\prime },L^{\prime })=Z(J,L),\qquad W^{\prime }(J^{\prime
},L^{\prime })=W(J,L).  \label{mapzw}
\end{equation}
The $L$-$J$ mixing of formula (\ref{bibip}) reflects the fact that a general
change of field variables mixes the elementary field with composite fields.
The transformations (\ref{bibip}) are associated with certain changes of
integration variables $\varphi ^{\prime }=$ $\varphi ^{\prime }(\varphi
,J,L) $ in the functional integral, combined with parameter redefinitions.

We say that the functional integral is written in the \textit{conventional
form} when the entire $J$-dependence is encoded in the term $\int J\varphi $
that appears in the exponent of the $Z$-integrand. Clearly, a non-linear
change of integration variables turns the functional integral into some
unconventional form. However, it was shown in ref. \cite{fieldcov} that the
conventional form can be recovered applying a nontrivial set of
manipulations. The renormalization of the theory in the new variables does
not need to be calculated anew. It can be derived from the renormalization
in the old variables applying the operations that switch the functional
integral back to the conventional form. The change of integration variables
undergoes its own renormalization, which is related to the renormalization
of composite fields.

In this paper we extend the investigation of ref. \cite{fieldcov} to the
generating functionals of one-particle irreducible diagrams. The usual
generating functional $\Gamma (\Phi ,L)$ is the Legendre transform of $W$
with respect to $J$. This kind of operation, however, must be treated with
caution, because it is not covariant. The first consequence of this fact is
that $\Gamma $ does not behave as a scalar under the field-transformation
law derived from its very definition as a $W$-Legendre transform. Yet, we
prove that there exists a corrected field-transformation law under which $%
\Gamma $ does behave as a scalar.

The second consequence is that there must exist a better generating
functional of one-particle irreducible diagrams, which does transform as
expected. We call it \textit{master functional} and denote it with $\Omega
(\Phi ,N)$. Roughly, it is the Legendre transform of $W$ with respect to
both $J$ and $L$. However, the naive Legendre transform with respect to $L$
does not exist, so we must first ``improve'' the $W$-functional in a
suitable way, and then make the Legendre transform of the improved $W$.

We said that the Legendre transform is not covariant under general source
redefinitions. Nevertheless, it is covariant under linear source
redefinitions. If we use the linear approach, where linear source
redefinitions encode the most general perturbative changes of field
variables, we do not lose generality. Approaches alternative to the linear
one have also been defined in ref. \cite{fieldcov}, but they are less
efficient for the purposes of this paper. For this reason here we mostly use
the so-called redundant linear approach, although we also include comments
on the other approaches. ``Redundant'' means that the basis $\{\mathcal{O}%
^{I}\}$ of composite fields is unrestricted. In particular, it contains also
descendants, composite fields proportional to the field equations, the
identity and the elementary field itself. In the redundant approach
divergent terms proportional to the field equations can still be subtracted
by means of field redefinitions, in the source-independent sector. Doing so
is useful, for example, to identify finite theories, whose divergences can
be subtracted by means of sole field redefinitions, and renormalizable
theories, whose divergences can be subtracted by means of field
redefinitions and redefinitions of a finite number of physical parameters.
In the source-dependent sector, instead, we take advantage of the redundancy
to simplify the formal structure as much as possible.

The master functional admits a convenient \textit{proper formulation}, where
the set of integrated fields $\varphi $ is extended to the \textit{proper
fields} $\varphi $-$N$, where $N$ are partners of the sources $L$. In the
proper formulation the classical action coincides with the classical limit $%
S_{N}$ of the master functional $\Omega $ and radiative corrections are
determined from the classical limit with the usual diagrammatic rules.
Moreover, the conventional form of the functional integral is manifestly
preserved during any change of field variables, including the BR map. In
this way, it is possible to work directly on $\Omega $ without referring to
its definition from $W$.

For definiteness, we work using the Euclidean notation and the dimensional
regularization, but no results depend on these choices. To simplify the
presentation, we imagine that the fields we are working with are bosonic,
but the arguments can be immediately generalized to include fermionic fields.

The paper is organized as follows. In section 2 we investigate how the
changes of field variables (\ref{bibip}) reflect inside the $\Gamma $%
-functional, and show that $\Gamma $ does not behave as a scalar under this
operation. We work out the correct field-transformation law under which $%
\Gamma $ does behave as a scalar. In section 3 we motivate the search for a
better generating functional of one-particle irreducible diagrams and
describe how to overcome the most basic difficulties. In section 4 we define
the master functional and investigate its main properties. We also calculate
it in an explicit example. In section 5 we study the perturbative changes of
field variables in the master functional, and apply them to the example of
section 4. In section 6 we study restrictions on the master functional, one
of which gives the $\Gamma $-functional itself. In section 7 we work out the
proper formulation, while in section 8 we study the renormalization of the
master functional. In section 9 we describe some generalizations obtained
``covariantizing'' the notion of Legendre transform. Section 10 contains the
conclusions, while in the appendix we recall a theorem used in the paper
about field redefinitions.

\section{Changes of field variables in the \texorpdfstring{$\Gamma$}{}%
-functional}

\label{unexpected}\setcounter{equation}{0}

In this section we study perturbative changes of field variables in the $%
\Gamma $-functional $\Gamma (\Phi ,L)=-W(J,L)+\int J\Phi $, where $\Phi
=\delta W/\delta J$ and $L$ are the sources coupled to the composite fields.
We first show that the transformation does not work as expected. Precisely,
under the transformation law derived from its very definition, $\Gamma $
does not behave as a scalar. This fact has an intuitive explanation. The
notion of one-particle irreducibility is not compatible with general field
redefinitions, because a non-linear change of field variables mixes
elementary fields with composite fields, therefore one-particle
irreducibility with many-particle irreducibility. We show that $\Gamma $
does transform as a scalar once we compose the expected change of field
variables with a further change of field variables.

Inside the functional $\Gamma (\Phi ,L)$ the change of variables that
follows from (\ref{bibip}), (\ref{mapzw}) and the definition of Legendre
transform reads 
\begin{equation}
\Phi ^{\prime }(\Phi ,L)=\frac{\delta W^{\prime }(J^{\prime },L^{\prime })}{%
\delta J^{\prime }}=\frac{\delta W(J,L)}{\delta J}+b_{I}\frac{\delta W(J,L)}{%
\delta L_{I}}=\Phi -b_{I}\frac{\delta \Gamma (\Phi ,L)}{\delta L_{I}}.
\label{fifip}
\end{equation}
We also have 
\begin{equation}
\frac{\delta \Gamma (\Phi ,L)}{\delta L_{I}}=-\frac{\delta W(J,L)}{\delta
L_{I}}=-\frac{\delta W^{\prime }(J,(L-bJ)z^{-1})}{\delta L_{I}}=-z^{-1}\frac{%
\delta W^{\prime }(J^{\prime },L^{\prime })}{\delta L_{I}^{\prime }}=z^{-1}%
\frac{\delta \Gamma ^{\prime }(\Phi ^{\prime },L^{\prime })}{\delta
L_{I}^{\prime }}.  \label{fillu}
\end{equation}

To visualize the change of variables more explicitly it is helpful to switch
composite fields off for a moment, setting $L=0$. Then the derivatives with
respect to the renormalized sources $L_{I}$ generate insertions of
renormalized composite fields $\mathcal{O}_{\mathrm{R}}^{I}$, so we get 
\[
\Phi ^{\prime }(\Phi ,0)=\Phi +b_{I}\left. \frac{\delta W}{\delta L_{I}}%
\right| _{L=0}=\langle \varphi +b_{I}\mathcal{O}_{\mathrm{R}}^{I}(\varphi
)\rangle _{L=0}\hspace{0.01in}. 
\]
Dropping also radiative corrections we see that the classical change of
variables is practically $\varphi ^{\prime }=\varphi +b_{I}\mathcal{O}%
_{c}^{I}(\varphi )$, where $\{\mathcal{O}_{c}^{I}\}$ is a basis of classical
composite fields, which coincide with the classical limits of the $\mathcal{O%
}_{\mathrm{R}}^{I}$s. Nevertheless, this result is only partially correct,
because the conditions $L=0$ switch composite fields off \textit{before} the
change of variables. After the change of variables we should impose $%
L^{\prime }=0$. It can be shown \cite{fieldcov} that when we take this fact
into account the correct classical change of variables becomes 
\[
\varphi ^{\prime }=\varphi +\tilde{b}_{I}\mathcal{O}_{c}^{I}(\varphi ), 
\]
where $\tilde{b}_{I}=b_{I}+\mathcal{O}(b^{2})$ is a calculable power series
in $b$.

Using (\ref{bibip}) and $J=\delta \Gamma /\delta \Phi $ we can express $%
L^{\prime }$ as a function of $\Phi $ and $L$. To express $\Phi $ and $L$ as
functions of $\Phi ^{\prime }$ and $L^{\prime }$, we can write 
\begin{eqnarray}
\Phi &=&\Phi ^{\prime }+b_{I}\frac{\delta \Gamma (\Phi ,L)}{\delta L_{I}}%
=\Phi ^{\prime }+bz^{-1}\frac{\delta \Gamma ^{\prime }(\Phi ^{\prime
},L^{\prime })}{\delta L_{I}^{\prime }}=\Phi (\Phi ^{\prime },L^{\prime }),
\label{accita} \\
L &=&L^{\prime }z+bJ^{\prime }=L^{\prime }z+b\frac{\delta \Gamma ^{\prime
}(\Phi ^{\prime },L^{\prime })}{\delta \Phi ^{\prime }}=L(\Phi ^{\prime
},L^{\prime }).  \label{cita}
\end{eqnarray}
Obviously, these relations, as well as (\ref{fifip}) and (\ref{fillu}), are
not linear and not even local.

Using (\ref{mapzw}), (\ref{fifip}) and the definitions of $\Gamma \ $and $%
\Gamma ^{\prime }$ we can work out the relation between the $\Gamma $%
-functionals. Keeping $\Phi $ and $L^{\prime }$ fixed and expanding in
powers of $b$ we find 
\begin{eqnarray}
\Gamma ^{\prime }(\Phi ^{\prime },L^{\prime }) &=&-W(J,L)+\int J\frac{\delta
W}{\delta J}(J,L)+\int b_{I}J\frac{\delta W}{\delta L_{I}}(J,L)=\Gamma (\Phi
,L)-\int b_{I}J\frac{\delta \Gamma (\Phi ,L)}{\delta L_{I}}  \nonumber \\
&=&\Gamma (\Phi ,L^{\prime }z)-\sum_{n=2}^{\infty }\frac{n-1}{n!}\int
J(bz^{-1})_{I_{1}}\cdots J(bz^{-1})_{I_{n}}\frac{\delta ^{n}\Gamma (\Phi
,L^{\prime }z)}{\delta L_{I_{1}}^{\prime }\cdots \delta L_{I_{n}}^{\prime }}.
\label{baty}
\end{eqnarray}
We could also expand in powers of $b$ keeping $\Phi $ and $L$ fixed,
instead, but it would give an equivalent result. The important thing is that
the same sources, in our case $L^{\prime }$, appear on the left- and
right-hand sides of (\ref{baty}). Then, setting $L^{\prime }=0$ we can
switch composite fields off both in the left- and right-hand sides of the
equation and compare $\Gamma ^{\prime }(\Phi ^{\prime },0)$ with $\Gamma
(\Phi ,0)$.

Formula (\ref{baty}) shows that the transformation rule we expect, $\Gamma
^{\prime }(\Phi ^{\prime },0)=\Gamma (\Phi ,0)$, does not hold, because
other terms appear on the right-hand side. Thus the change of variables in
the $\Gamma $-functional does not work as expected.

Nevertheless, we can show that the extra terms that appear in the last line
of (\ref{baty}) can be reabsorbed inside a further change of variables.
Recalling that 
\begin{equation}
J=J^{\prime }=\frac{\delta \Gamma ^{\prime }(\Phi ^{\prime },L^{\prime })}{%
\delta \Phi ^{\prime }},  \label{jj}
\end{equation}
we can manipulate the last line of (\ref{baty}) and get 
\begin{equation}
\Gamma (\Phi ,L^{\prime }z)=\Gamma ^{\prime }(\Phi ^{\prime },L^{\prime
})+\sum_{n=2}^{\infty }\frac{n-1}{n!}\int \frac{\delta \Gamma ^{\prime }}{%
\delta \Phi ^{\prime }}(bz^{-1})_{I_{1}}\cdots \frac{\delta \Gamma ^{\prime }%
}{\delta \Phi ^{\prime }}(bz^{-1})_{I_{n}}\left. \frac{\delta ^{n}\Gamma
(\Phi ,L^{\prime }z)}{\delta L_{I_{1}}^{\prime }\cdots \delta
L_{I_{n}}^{\prime }}\right| _{\Phi =\Phi (\Phi ^{\prime },L^{\prime })}.
\label{coeffo}
\end{equation}
Observe that the corrections to $\Gamma ^{\prime }$ on the right-hand side
are at least quadratic in the field equations of $\Gamma ^{\prime }$. Thanks
to this fact, we can apply a theorem of ref. \cite{acaus}, recalled in the
appendix. That theorem ensures that the corrections of (\ref{coeffo}) can be
reabsorbed inside $\Gamma ^{\prime }$ by means of a further (still
non-local) change of variables $\tilde{\Phi}(\Phi ^{\prime },L^{\prime })$.
The change of variables is encoded in formulas (\ref{redef}) and (\ref
{genfor}) of the appendix, while the structure of the action and the
transformation law are given by formulas (\ref{ayu}) and (\ref{equa}). We
obtain 
\begin{equation}
\Gamma (\Phi ,L^{\prime }z)=\Gamma ^{\prime }(\tilde{\Phi}(\Phi ^{\prime
},L^{\prime }),L^{\prime }).  \label{f1}
\end{equation}
Defining $\Gamma ^{\prime \prime }(X,\tilde{L})=\Gamma ^{\prime
}(X,L^{\prime })$ and the corrected change of variables 
\begin{equation}
\Phi ^{\prime \prime }(\Phi ,\tilde{L})\equiv \tilde{\Phi}(\Phi ^{\prime
}(\Phi ,L^{\prime }),L^{\prime }),  \label{nonlocchv}
\end{equation}
where $\tilde{L}=L^{\prime }z$ and $\Phi ^{\prime }(\Phi ,L^{\prime })$ is
obtained inverting (\ref{accita}), we get 
\begin{equation}
\Gamma (\Phi ,\tilde{L})=\Gamma ^{\prime \prime }(\Phi ^{\prime \prime },%
\tilde{L}),  \label{f2}
\end{equation}
namely the $\Gamma $-functional transforms as a scalar under the corrected
transformation law. Now we can forget about the origin of the sources $%
\tilde{L}$ and just pay attention to the fact that they are the same on both
sides of the equation.

In particular, when composite fields are switched off ($\tilde{L}=0$) the
transformation law for the $\Gamma $-functional reads 
\[
\Gamma (\Phi )=\Gamma (\Phi ,0)=\Gamma ^{\prime \prime }(\Phi ^{\prime
\prime }(\Phi ,0),0)=\Gamma ^{\prime \prime }(\Phi ^{\prime \prime }), 
\]
where in the last expression it is understood that $\Phi ^{\prime \prime }$
is $\Phi ^{\prime \prime }(\Phi ,0)$.

Now we prove that (\ref{nonlocchv}) is the correct change of field variables
for the functional $\Gamma $. To do so we must carefully analyze the
structure of the $\Gamma $-functional and the properties of its changes of
field variables. Non-local field redefinitions are tricky, because they give
us an enormous freedom and can even relate theories that are not physically
equivalent to each other, if we do not apply them correctly. Acceptable
changes of field variables are only those that do relate physically
equivalent theories.

The $\Gamma (\Phi ,L)$-functional can be decomposed into the sum of a local
tree-level action $S_{cL}(\Phi ,L)$, which coincides with the classical
action, plus (non-local) radiative corrections $\hbar \Gamma _{\text{non-loc}%
} $. The radiative corrections are determined by the tree-level action
itself. Precisely, $\hbar \Gamma _{\text{non-loc}}$ collects the
one-particle irreducible diagrams that are constructed with the vertices and
propagators determined by $S_{cL}$, multiplied by appropriate coefficients.
We write 
\begin{equation}
\Gamma (\Phi ,L)=S_{cL}(\Phi ,L)+\hbar \Gamma _{\text{non-loc}}(\Phi ,L).
\label{gastru}
\end{equation}

A non-local field redefinition $\Phi ^{\prime \prime }=\Phi ^{\prime \prime
}(\Phi ,\tilde{L})$ maps physically equivalent theories when it is a
perturbative field redefinition at the tree level and $\Gamma ^{\prime
\prime }(\Phi ^{\prime \prime },\tilde{L})=\Gamma (\Phi ,\tilde{L})$ has a
structure analogous to (\ref{gastru}). We can decompose it as 
\[
\Phi ^{\prime \prime }=\Phi ^{\prime \prime }(\Phi ,\tilde{L})=\Phi
_{c}^{\prime \prime }(\Phi ,\tilde{L})+\hbar \Phi _{\text{non-loc}}^{\prime
\prime }(\Phi ,\tilde{L}), 
\]
where $\Phi _{c}^{\prime \prime }(\Phi ,\tilde{L})$ is local. Moreover, the
field redefinition must be such that the new non-local radiative corrections 
$\hbar \Gamma _{\text{non-loc}}^{\prime \prime }$ collect the one-particle
irreducible diagrams determined by the new classical action $S_{cL}^{\prime
\prime }$, multiplied by the correct coefficients.

Now we prove that the non-local change of variables (\ref{nonlocchv})
satisfies these requirements. First, let us recall the form of the classical
action $S_{cL}$ in the redundant linear approach, because we are going to
use it in the proof. It is given by 
\begin{equation}
S_{cL}(\varphi ,L)=S_{c}(\varphi )-\int L_{I}\mathcal{O}_{c}^{I}(\varphi
)-\int \tau _{vI}\mathcal{N}^{v}(L)\mathcal{O}_{c}^{I}(\varphi ),
\label{bibop}
\end{equation}
where $\mathcal{N}^{v}(L)$ is a basis of independent local monomials that
can be constructed with the sources $L$ and their derivatives, and are at
least quadratic in $L$, while the $\tau _{vI}$s are constants. The reason
why composite fields are multiplied by the most general $\mathcal{O}(L)$%
-structure is that doing so it is possible to linearize also the BR map,
which can be expressed as a source redefinition of the form (\ref{bibip})
combined with parameter redefinitions.

Now, let us consider the map $\Phi ^{\prime }(\Phi ,L^{\prime })$. We can
work it out inverting formula (\ref{accita}) perturbatively in $b$. If we
are just interested in the tree-level contributions to $\Phi ^{\prime }(\Phi
,L^{\prime })$ we can use (\ref{fillu}) and (\ref{jj})\ to replace $\delta
\Gamma ^{\prime }/\delta L_{I}^{\prime }$ and $\delta \Gamma ^{\prime
}/\delta \Phi ^{\prime }$ with $\delta \Gamma /\delta L_{I}$ and $\delta
\Gamma /\delta \Phi $ in (\ref{accita}) and (\ref{cita}). Then we can
replace $\Gamma $ with $S_{cL}$, given by (\ref{bibop}), then iterate (\ref
{cita}) to express $L$ as a function of $\Phi $ and $L^{\prime }$, insert
the result in (\ref{accita}), and finally invert (\ref{accita}). Clearly,
the result is a perturbative field redefinition. We conclude that $\Phi
^{\prime }=\Phi ^{\prime }(\Phi ,L^{\prime })$ is a perturbative field
redefinition plus radiative corrections. Next, consider the map $\tilde{\Phi}%
(\Phi ^{\prime },L^{\prime })$. Observe that, expressed in the variables $%
\Phi $-$L^{\prime }z$ the coefficients of the $\delta \Gamma ^{\prime
}/\delta \Phi ^{\prime }$-powers in (\ref{coeffo}) are local at the tree
level, and at higher orders they involve only one-particle irreducible
diagrams with multiple composite-field insertions. These properties hold
even after expressing $\Phi $ as a function of $\Phi ^{\prime }$ and $%
L^{\prime }$, which is done using (\ref{accita}). Thus, using formulas (\ref
{redef}) and (\ref{genfor}), we see that $\tilde{\Phi}^{\prime }(\Phi
^{\prime },L^{\prime })$ shares the same properties. Composing this
transformation with $\Phi ^{\prime }=\Phi ^{\prime }(\Phi ,L^{\prime })$, we
find that $\Phi ^{\prime \prime }(\Phi ,\tilde{L})$ is the sum of a
tree-level perturbative field redefinition plus radiative corrections that
involve only one-particle irreducible diagrams, as we wished to prove.

The second requirement, that the radiative corrections are determined by the
tree-level action with the usual diagrammatic rules, is also satisfied.
Indeed, definition $\Gamma ^{\prime \prime }(X,\tilde{L})=\Gamma ^{\prime
}(X,L^{\prime })$ tells us that the transformed functional $\Gamma ^{\prime
\prime }(\Phi ^{\prime \prime },\tilde{L})$ is just the functional $\Gamma
^{\prime }(\Phi ^{\prime },L^{\prime })$ with $\Phi ^{\prime }$ replaced by $%
\Phi ^{\prime \prime }$. We know that $\Gamma ^{\prime }$ has the correct
structure, which is the primed version of (\ref{gastru}), therefore $\Gamma
^{\prime \prime }$ also has the correct structure.

We conclude that (\ref{nonlocchv}) is an acceptable change of variables for
the $\Gamma $-functional, which behaves as a scalar.

Observe that all non-localities involved in the change of field variables
are those typical of one-particle irreducible diagrams. Nowhere the
non-localities typical of the $W$-functional (such as propagators with
external momenta) enter the game. The change of variables itself is
one-particle irreducible.

So far we have used the linear redundant approach, taking (\ref{bibop}) as
the classical action and assuming that the most general change of field
variables is encoded in the source redefinitions (\ref{bibip}).
Nevertheless, the argument can be easily generalized to the essential
approach and the other non-linear approaches studied in ref. \cite{fieldcov}%
. In the essential approach, which is inspired by the classification of
couplings made in ref. \cite{weinberg}, we work with a basis of composite
fields that does not contain descendants (i.e. derivatives of other
composite fields) and objects proportional to the field equations. Then some
changes of field variables, for example those appearing in the BR map,
require to make non-linear source transformations in $W$. Something similar
occurs with the other approaches of ref. \cite{fieldcov}. Consider the most
general perturbatively local finite redefinitions 
\begin{equation}
L_{I}^{\prime }=L_{I}^{\prime }(J,L)=L_{I}^{\prime }+\mathcal{O}(b),\qquad
J^{\prime }=J,  \label{bu}
\end{equation}
that can be expanded in powers of some parameters $b$ and satisfy the
initial conditions $L_{I}^{\prime }(0,0)=0$. We recall that every
transformed functional $W^{\prime }(J^{\prime },L^{\prime })=W(J,L)$,
obtained applying (\ref{bu}), is the $W$-functional that we would calculate
in some transformed field-variable frame. The transformed fields can be
worked out applying the procedure explained in ref. \cite{fieldcov} to
recover the conventional form of the functional integral, which is spoiled
by any nontrivial $J$-dependence contained in $L^{\prime }(J,L)$.

At the level of the $\Gamma $-functional, the expected field transformation
reads 
\begin{equation}
\Phi ^{\prime }(\Phi ,L)=\Phi -\int \frac{\delta L_{I}(J^{\prime },L^{\prime
})}{\delta J^{\prime }}\frac{\delta \Gamma (\Phi ,L)}{\delta L_{I}}.
\label{buo}
\end{equation}
Expanding in powers of $J$ we can write 
\begin{equation}
\Gamma ^{\prime }(\Phi ^{\prime },L^{\prime })=\Gamma (\Phi ,L)-\int
J^{\prime }\frac{\delta L_{I}(J^{\prime },L^{\prime })}{\delta J^{\prime }}%
\frac{\delta \Gamma (\Phi ,L)}{\delta L_{I}}=\Gamma (\Phi ,L(L^{\prime
}))-\int J^{\prime }\mathcal{M}(\Phi ,L^{\prime })J^{\prime },  \label{bii}
\end{equation}
where $L(L^{\prime })=L(0,L^{\prime })$ and $\mathcal{M}(\Phi ,L^{\prime })$
is an order $b^{2}$-sum of a tree-level local functional plus one-particle
irreducible radiative corrections. Then we use (\ref{bu}) and (\ref{buo}) to
express $\Phi $ as a function of $\Phi ^{\prime }$ and $L^{\prime }$ inside $%
\mathcal{M}$, move the last term (\ref{bii}) to the left-hand side, realize
that the correction to $\Gamma ^{\prime }$ is quadratically proportional to
the $\Gamma ^{\prime }$-field equations, and reabsorb such a correction into
a further change of variables $\tilde{\Phi}(\Phi ^{\prime })$, applying the
theorem recalled in the appendix. All arguments proceed as above, with
straightforward modifications, and lead us to conclude that the final change
of variables (\ref{nonlocchv}) is correct, because it preserves the
structure (\ref{gastru}) of the $\Gamma $-functional, which expresses $%
\Gamma $ as the sum of a local function plus one-particle irreducible
radiative corrections, determined by the tree-level part with the usual
diagrammatic rules.

Summarizing, the final change of variables for $\Gamma $ is not the one
inherited by the very definition of $\Gamma $ as the Legendre transform of $%
W $, namely (\ref{fifip}) or (\ref{buo}), but instead it is (\ref{nonlocchv}%
). Nevertheless, the result we have found proves that a correct change of
field variables for $\Gamma $ does exist. We just need to bear in mind that
it is not the expected one.

In the next sections we show that there exists a better functional that
still collects one-particle irreducible diagrams, but also transforms as
expected, and very simply (that is to say linearly, in the linear approach),
under arbitrary changes of field variables.

\section{Master functional: motivation and introductory observations}

\setcounter{equation}{0}

A change of field variables in the functionals $Z$ and $W$ is a redefinition
of the sources $J$ and $L$. Although the functionals $Z$ and $W$ are
non-local, the $J$- and $L$-redefinitions must be local, since the exponent
of the $Z$-integrand 
\begin{equation}
Z(J,L)=\int [\mathrm{d}\varphi ]\exp \left( -S_{L}(\varphi ,L)+\int J\varphi
\right)  \label{uy0}
\end{equation}
must remain local. In the linear approach, the $J$- and $L$-redefinitions
are local \textit{and} linear. Moreover, since we include the elementary
field in the set of composite fields, we can work in a framework where $J$
is unmodified and the entire transformation is encoded in the $L$%
-redefinition, as shown in (\ref{bibip}).

On the other hand, we have observed that under changes of field variables
the generating functional $\Gamma (\Phi ,L)$ of one-particle irreducible
diagrams does not transform as expected from its very definition as the
Legendre transform of $W$. We have been able to find a more involved change
of variables that compensates for this fact and that is satisfactory for
most purposes. The complete field redefinition is itself non-local. This is
not surprising, because $\Gamma $ is a non-local functional.

Nevertheless, since the source redefinitions (\ref{bibip}) for $Z$ and $W$
are local, and linear in the linear approach, we are tempted to think that
there should exist a better generating functional $\Omega $ of one-particle
irreducible diagrams that works similarly, namely such that the most general
field transformations can be expressed locally, and linearly in the linear
approach. Moreover, the transformations should be the ones following from
the very definition of $\Omega $. In this section we collect a number of
remarks that help us identify the desired generating functional.

Intuitively, the problem of $\Gamma $ is that a non-linear change of
variables mixes the elementary field with composite fields, therefore
one-particle irreducibility with many-particle irreducibility. This argument
suggests that maybe we should work with many-particle irreducible generating
functionals \cite{manyirr}. Recall, however, that those generating
functionals are defined coupling non-local sources $K_{n}(x_{1},\ldots
,x_{n})$ with strings $\varphi (x_{1})\cdots \varphi (x_{n})$ of
elementary-field insertions located at distinct points, which are non-local
composite fields. When we want to study local changes of variables we need
to shift the sources $K$ by local terms proportional to $J$. For example,
the shift 
\begin{equation}
K_{2}(x,y)\rightarrow K_{2}(x,y)+b\delta (x-y)J(x),  \label{kshi}
\end{equation}
allows us to study the change of variables $\varphi \rightarrow \varphi
+b\varphi ^{2}$. However, non-local sources do not capture the
renormalization of local composite fields. Thus, the local shift of (\ref
{kshi}) causes the appearance of new divergences, those associated with the
composite field $\varphi ^{2}$, which need to be calculated anew in this
approach. For this reason, we do not pursue the use of generating
functionals of many-particle irreducible diagrams and look for a different
solution.

Since a change of variables mixes the elementary field with (local)
composite fields, it sounds natural to treat all of them on the same
footing. This suggests to define a functional $\Omega (\Phi ,N)$ as the
Legendre transform of the functional $W(J,L)$ with respect to all sources $J$
and $L$, not just with respect to $J$. However, the Legendre transform of $W$
with respect to the sources $L$ does not exist, in general. The two-point
functions of composite fields in momentum space form a matrix $G^{IJ}$ that
is not invertible, due to some sort of ``gauge'' symmetries obeyed by the
sources $L$.

Before moving forward, let us illustrate this important point in more
detail. Composite fields proportional to the field equations give zero or a
contact term, when they are inserted in a two-point function. On the other
hand, descendants give two-point functions proportional to the ones of their
primaries, so if the matrix $G^{IJ}$ contains both primaries and descendants
it is degenerate. We might think that in the essential approach, where
descendants and composite fields proportional to the field equations are not
contained in the basis of composite fields, $G^{IJ}$ is invertible. This is
not true, however.

Consider a free massless scalar field $\varphi $ in Euclidean space and the
composite fields $\mathcal{O}^{J}=\varphi ^{J}/J!$. The two-point functions $%
G^{IJ}=\langle \mathcal{O}^{I}\mathcal{O}^{J}\rangle _{L=0}$ can be easily
calculated in momentum space integrating one loop after another. The result
is, using the dimensional-regularization technique, 
\begin{equation}
\langle \mathcal{O}^{I}(k)\ \mathcal{O}^{J}(-k)\rangle =\frac{\delta ^{IJ}}{%
J!}\frac{\Gamma \left( J-D\frac{J-1}{2}\right) \left[ \Gamma \left( \frac{D}{%
2}-1\right) \right] ^{J}}{(4\pi )^{(J-1)D/2}\Gamma \left( \frac{JD}{2}%
-J\right) }(k^{2})^{(J-1)D/2-J},  \label{buri}
\end{equation}
where $D=4-\varepsilon $ is the continued spacetime dimension. Subtracting
the divergent part at coincident points, we find 
\begin{equation}
\left. \langle \mathcal{O}^{I}(k)\ \mathcal{O}^{J}(-k)\rangle \right| _{%
\text{finite}}=\delta ^{IJ}\frac{(-1)^{J-1}(k^{2})^{J-2}(\ln k^{2}+\text{%
constant})}{(4\pi )^{2J-2}J!(J-1)!(J-2)!}.  \label{byji}
\end{equation}

The matrix (\ref{byji}) is a diagonal block of $G^{IJ}$ and is invertible.
Nevertheless, observe that it would be problematic to use its inverse. The
reason is that the first nontrivial contributions to (\ref{byji}) are
one-loop, so its inverse introduces negative powers of $\hbar $. At the bare
level, the matrix is even divergent, so its inverse introduces objects of
order $\varepsilon $, very difficult to handle.

Now we calculate the $G^{IJ}$-block made of the composite fields $\{\mathcal{%
O}^{1},\mathcal{O}^{2}\}=\{(1/2)\varphi ^{2},(1/2)\varphi \partial _{\mu
}\partial _{\nu }\varphi \}$. We find, in momentum space, 
\begin{eqnarray}
\langle \varphi ^{2}\ \varphi ^{2}\rangle &=&-\frac{\ln (k^{2}/\mu ^{2})}{%
32\pi ^{2}},\qquad \langle \varphi ^{2}\ \mathcal{O}_{M}\rangle =-\frac{\ln
(k^{2}/\mu ^{2})}{96\pi ^{2}}M_{\mu \nu }k_{\mu }k_{\nu },  \nonumber \\
\langle \mathcal{O}_{M}\ \mathcal{O}_{M}\rangle &=&-\frac{\ln (k^{2}/\mu
^{2})}{3840\pi ^{2}}\left( (k^{2})^{2}M_{\mu \nu }^{2}-2k^{2}(M_{\mu \nu
}k_{\nu })^{2}+14(M_{\mu \nu }k_{\mu }k_{\nu })^{2}\right) ,  \label{block}
\end{eqnarray}
where $M_{\mu \nu }$ is a constant traceless matrix and $\mathcal{O}%
_{M}=(M_{\mu \nu }/2)\varphi \partial _{\mu }\partial _{\nu }\varphi $. This 
$G^{IJ}$-block is not invertible. A quick way to prove this statement is to
check that the vector $\{(k^{2})^{2},\delta _{\mu \nu }k^{2}-4k_{\mu }k_{\nu
}\}$ is a null vector.

We can interpret this singularity as the consequence of a gauge symmetry.
Although $\varphi \partial _{\mu }\partial _{\nu }\varphi $ is not a
descendant of $\varphi ^{2}$, the two composite fields $\varphi \partial
_{\mu }\partial _{\nu }\varphi $ and $\varphi ^{2}$ have a descendant in
common, up to terms proportional to the field equations. Indeed, 
\[
\partial _{\nu }(\varphi \partial _{\mu }\partial _{\nu }\varphi )=\frac{1}{4%
}\partial _{\mu }\Box (\varphi ^{2})+\varphi (\partial _{\mu }\Box \varphi )-%
\frac{1}{2}\partial _{\mu }(\varphi \Box \varphi ). 
\]
The action 
\begin{equation}
S_{L}=\frac{1}{2}\int (\partial \varphi )^{2}-\frac{1}{2}\int L\varphi ^{2}-%
\frac{1}{2}\int L_{\mu \nu }(\varphi \partial _{\mu }\partial _{\nu }\varphi
)  \label{thisfact}
\end{equation}
is invariant with respect to the infinitesimal ``gauge'' transformation 
\begin{equation}
\delta \varphi =-\partial \cdot (\varphi \ell )+\frac{1}{2}\varphi (\partial
\cdot \ell ),\qquad \delta L_{\mu \nu }=\partial _{\mu }\ell _{\nu
}+\partial _{\nu }\ell _{\mu },\qquad \delta L=-\frac{1}{2}\Box (\partial
\cdot \ell ),  \label{gaesym}
\end{equation}
to the lowest order in $L$, where $\ell _{\mu }$ are arbitrary functions.
This is why the block cannot be invertible.

The symmetry (\ref{gaesym}) can be extended to the complete action 
\[
S_{L}=\frac{1}{2}\int (\partial \varphi )^{2}-\int L_{I}\mathcal{O}^{I}, 
\]
assuming that $\{\mathcal{O}^{I}\}$ is the basis of composite fields. We
must cancel the terms 
\begin{equation}
-\int L_{I}\frac{\delta \mathcal{O}^{I}}{\delta \varphi }\delta \varphi ,
\label{remno}
\end{equation}
which can be done as follows. Expanding (\ref{remno}) in the basis $\{%
\mathcal{O}^{I}\}$, we can write (\ref{remno}) as 
\[
-\int P_{I}(L,\ell )\mathcal{O}^{I}(\varphi ), 
\]
where $P_{I}(L,\ell )$ are bilinear local functions of $L$ and $\ell $, or
their derivatives. Then to reabsorb (\ref{remno}) it is sufficient to
correct the $\delta L_{I}$-transformations of (\ref{gaesym}) as 
\[
\delta L_{I}\rightarrow \delta L_{I}-P_{I}(L,\ell ). 
\]
Note that the $L$-transformations remain $\varphi $-independent, as it must
be, otherwise it would be impossible to apply them inside the functional
integral.

Clearly, similar arguments can be used to relate most of the composite
fields containing derivatives. We learn that the renormalized two-point
functions $G^{IJ}$, which are equal to $\langle \mathcal{O}_{\mathrm{R}}^{I}%
\mathcal{O}_{\mathrm{R}}^{J}\rangle _{L=0}$ plus counterterms taking care of
coinciding points, do \textit{not} form an invertible matrix in momentum
space, not even if the set $\{\mathcal{O}^{I}\}$ is restricted to the
essential fields. Thus, the Legendre transform of the $\Gamma $-functional
with respect to the sources $L$ does not exist, in general.

At the same time, we learn that this problem is due to the presence of a
special class of gauge symmetries. One way to solve it is to gauge-fix those
gauge symmetries. However, since the sources $L$ are just formal tools, we
do not need to worry about the propagation of unphysical ``$L$-degrees of
freedom''. Therefore, more simply, we can just break the symmetries (\ref
{gaesym}) explicitly.

We need to choose the most convenient symmetry-breaking term. We can show
that the unique term that is compatible with all properties we need (some of
which we have not mentioned, yet) is 
\begin{equation}
T(L)=\frac{1}{2}\int L_{I}(A^{-1})^{IJ}L_{J},  \label{improfin}
\end{equation}
where $A$ is a constant invertible matrix. We call (\ref{improfin}) \textit{%
improvement term}. It must be included in $-S_{L}$, and therefore also $%
W(J,L)$, if it is not already present. It provides otherwise missing
tree-level quadratic contributions for the $L$-sector. If we proceed in the
way explained below, this trick is enough to make the $W$-Legendre transform
with respect to the sources $L$ well-defined.

Now we must face another key problem: the Legendre transform is not a
covariant operation. Given a function $f(x^{\mu })$, define $y_{\mu }=%
\mathrm{d}f/\mathrm{d}x^{\mu }$ and the Legendre transform $%
g(y)=-f(x(y))+x^{\mu }(y)y_{\mu }$. Consider a general change of coordinates 
$x^{\prime }=x^{\prime }(x)$ and study how it reflects from $f$ to $g$. To
do this, it is useful to write $g$ as a function of $x$: 
\begin{equation}
g=-f(x)+x^{\mu }\frac{\mathrm{d}f}{\mathrm{d}x^{\mu }}.  \label{pegepo}
\end{equation}
If $f$ transforms as a scalar, then $\mathrm{d}f/\mathrm{d}x^{\mu }$
transforms as a vector. However, $x^{\mu }$ does not transform as a vector,
so $g$ is not a scalar.

There is one exception: the Legendre transform $g$ does behave as a scalar
when the change of coordinates $x^{\prime }=x^{\prime }(x)$ is linear. If we
use the linear approach, where all changes of field variables can be
expressed as linear transformations of $L$ and $J$, we can define a
completely invariant $\Omega $.

This is encouraging, yet still not enough for our purposes. The main virtue
of the functional $\Gamma $ is that its diagrams obey the theorem of
locality of counterterms. Because of this, it is relatively easy to have
control on renormalization working on $\Gamma $. It is more difficult
working, for example, directly on $W$, where local divergences can be
multiplied by propagators and generate non-local divergent expressions.

Thus, the functional $\Omega $ must be a collection of one-particle
irreducible diagrams. Better, it must be a collection of one-particle
irreducible diagrams in all variable frames. To achieve this result it is
sufficient to require that the ``propagators'' of the sources $L$ be equal
to the identity. In this way, $L$-insertions are glued together at the same
point and no $W$-type of non-localities are generated. More details on this
issue are given in the next section.

The desired type of $L$-propagators are given by the improvement term (\ref
{improfin}), therefore it is sufficient to state that all other $\mathcal{O}%
(L^{2})$-terms belonging to the $L$-sector must be treated perturbatively
with respect to (\ref{improfin}). We show below that it is consistent to do
so.

In the end, we are able to build a functional $\Omega $ that meets our
requirements. It is invariant with respect to the most general changes of
field variables, it is one-particle irreducible in all field-reference
frames and it obeys the theorem of locality of counterterms. Moreover, it
contains all pieces of information we need, since we can always reconstruct $%
W$ and $Z$ (and also $\Gamma $) from $\Omega $. Finally, we can renormalize
the theory working directly on $\Omega $ instead of $\Gamma $.

We think that the functional $\Omega $ can play a key role in the general
field-covariant approach to quantum field theory. This is the reason why we
call it the master functional.

We have already noted that the linear approach is very convenient for our
purposes, because there all changes of field variables, including the BR
map, are described by linear source redefinitions, which are transparent to
the Legendre transform. Moreover, the linear approach provides the
improvement term (\ref{improfin}) naturally, because it is contained inside
the terms $\tau _{v0}\int \mathcal{N}^{v}(L)$ that multiply the identity
operator in the classical extended action $S_{cL}$ (\ref{bibop}). In case
that term is not already present, we just add it. Actually, for future use
it is better to shift $\tau _{v0}\int \mathcal{N}^{v}(L)$ by $T(L)$, even if
this operation may introduce some redundancy.

The classical action (\ref{bibop}) is now turned into 
\begin{equation}
S_{cL}(\varphi ,\lambda ,L)=S_{c}(\varphi ,\lambda )-\int L_{I}\mathcal{O}%
_{c}^{I}(\varphi ,\lambda )-T(L)-\int \tau _{vI}\mathcal{N}^{v}(L,\lambda )%
\mathcal{O}_{c}^{I}(\varphi ,\lambda ),  \label{illo}
\end{equation}
where $\lambda $ are the masses, the coupling constants and all other
parameters of the theory. The bare action is formally identical, with bare
quantities replacing classical quantities: $S_{L\mathrm{B}}(\varphi _{%
\mathrm{B}},\lambda _{\mathrm{B}},L_{\mathrm{B}})=S_{cL}(\varphi _{\mathrm{B}%
},\lambda _{\mathrm{B}},L_{\mathrm{B}})$, $S_{\mathrm{B}}(\varphi _{\mathrm{B%
}},\lambda _{\mathrm{B}})=S_{c}(\varphi _{\mathrm{B}},\lambda _{\mathrm{B}})$%
, $\mathcal{O}_{\mathrm{B}}^{I}(\varphi _{\mathrm{B}},\lambda _{\mathrm{B}})=%
\mathcal{O}_{c}^{I}(\varphi _{\mathrm{B}},\lambda _{\mathrm{B}})$. Finally,
let 
\[
\varphi _{\mathrm{B}}=\varphi _{\mathrm{B}}(\varphi ,\lambda ,\mu ),\qquad
\lambda _{\mathrm{B}}=\lambda _{\mathrm{B}}(\lambda ,\mu ), 
\]
be the relation between bare and renormalized fields and couplings when
composite fields are switched off. The renormalized action reads \cite
{fieldcov} 
\begin{equation}
S_{L}(\varphi ,\lambda ,\mu ,L)=S(\varphi ,\lambda ,\mu )-T(L)-\int \left(
L_{I}+\hat{\tau}_{vI}\mathcal{N}^{v}(L,\lambda ,\mu )\right) \mathcal{O}_{%
\mathrm{R}}^{I}(\varphi ,\lambda ,\mu ),  \label{sl}
\end{equation}
where $\hat{\tau}=\tau $ plus counterterms, $S$ is the renormalized action
at $L=0$ and $\mathcal{O}_{\mathrm{R}}^{I}$ are the renormalized composite
fields. We have 
\[
S(\varphi ,\lambda ,\mu )=S_{\mathrm{B}}(\varphi _{\mathrm{B}},\lambda _{%
\mathrm{B}}),\qquad \mathcal{O}_{\mathrm{R}}^{I}=(Z^{-1})_{J}^{I}\mathcal{O}%
^{J}(\varphi ,\lambda ,\mu )=(Z^{-1})_{J}^{I}\mathcal{O}_{\mathrm{B}%
}^{I}(\varphi _{\mathrm{B}},\lambda _{\mathrm{B}}), 
\]
where $Z_{J}^{I}$ is the matrix of renormalization constants for the
composite fields. If counterterms of type $T(L)$ are necessary, we include
them in $\mathbf{\tau }_{v0}\int \mathcal{N}^{v}$ and keep $T(L)$
unrenormalized.

Next, we state that when we make the Legendre transform with respect to the
sources $L$, $\tau _{v0}\int \mathcal{N}^{v}$ must be treated perturbatively
with respect to (\ref{improfin}). This is achieved as follows. In ref. \cite
{fieldcov} it was shown that the perturbative expansion is well organized if
we assume 
\begin{equation}
\lambda _{n_{l}}=\mathcal{O}(\delta ^{n_{l}-2}),\qquad L_{I}=\mathcal{O}%
(\delta ^{n_{I}-2}),\qquad A_{IJ}=\mathcal{O}(\delta
^{n_{I}+n_{J}-2}),\qquad \tau _{vI}=\mathcal{O}(\delta ^{n_{I}-n_{v}-2}),
\label{assigne0}
\end{equation}
where $\delta $ is some reference parameter $\ll 1$. Here $\lambda _{n_{l}}$
is the coupling, or product of couplings, that multiplies a monomial with $%
n_{l}$ $\varphi $-legs, $n_{I}$ is such that $\mathcal{O}^{I}(\varphi \delta
^{-1},\lambda _{l}\delta ^{n_{l}-2})=\delta ^{-n_{I}}\mathcal{O}^{I}(\varphi
,\lambda )$ and $n_{v}$ is the $\delta $-degree of $\mathcal{N}^{v}(L)$.
With these assignments all radiative corrections carry an extra factor $%
\delta ^{2\ell }$, where $\ell $ is the number of loops. However, for the
present purposes we need to slightly modify the assignment (\ref{assigne0}),
in a way that makes $A^{-1}$ more important than the $\tau $s and does not
affect the statements derived so far. For example we can assume that $\tau
_{vI}$ is $\mathcal{O}(\delta ^{n_{I}-n_{v}-1})$, while $A_{IJ}$ remains $%
\mathcal{O}(\delta ^{n_{I}+n_{J}-2})$. In this way all $\tau _{vI}$s remain
leading with respect to their radiative corrections, so the assignment
modification is consistent with our previous arguments. Summarizing, the
perturbative expansion is properly organized assuming 
\begin{equation}
\lambda _{n_{l}}=\mathcal{O}(\delta ^{n_{l}-2}),\qquad L_{I}=\mathcal{O}%
(\delta ^{n_{I}-2}),\qquad A_{IJ}=\mathcal{O}(\delta
^{n_{I}+n_{J}-2}),\qquad \tau _{vI}=\mathcal{O}(\delta ^{n_{I}-n_{v}-1}),
\label{assigne1}
\end{equation}
instead of (\ref{assigne0}). Consistently with (\ref{assigne1}), we also
have $J=\mathcal{O}(\delta ^{-1})$, since both $J$ and $L_{1}$ are sources
for the elementary field. These assignments are easy to remember, because if
we rescale every object by a factor $\delta ^{n}$, where $n$ it its $\delta $%
-degree, and in addition rescale $\varphi $ by $1/\delta $, then the action $%
S_{L}$ rescales as 
\[
S_{L}\rightarrow \frac{1}{\delta ^{2}}\bar{S}_{L}, 
\]
where $\bar{S}_{L}$ has a factor $\delta $ for each $\tau $ and a factor $%
\delta ^{2}$ for each loop, but is $\delta $-independent everywhere else.

Before concluding this section, let us explain why (\ref{improfin}) is
unique for our purposes. Under a Legendre transform the coefficients of
quadratic terms are turned into their reciprocals. If, for example, (\ref
{improfin}) were replaced with $L$-quadratic terms containing polynomials in
derivatives, the $L$-propagators would be non-local. Then the master
functional would contain unphysical poles, one-particle irreducibility would
be destroyed and the theorem of locality of counterterms would be difficult
to apply. To avoid all this, the $L$-propagators must be local. Now, assume
that (\ref{improfin}) is replaced with a non-local improvement term, such
that the $L$-propagators are still local. A non-local improvement term of
this type is acceptable inside $W$, which is non-local, but not acceptable
in the exponent of the $Z$-integrand, which must be local. However, in these
two places the improvement term is just the same. We conclude that both the
improvement term and the $L$-propagators derived from it should be local,
which leaves just (\ref{improfin}).

\section{Master functional:\ definition and basic properties}

\setcounter{equation}{0}

Now we are ready to define the master functional and study its structure. As
said, we use the redundant linear approach. Moreover, we work at the
renormalized level, because the arguments extend to bare quantities with
little modifications. Let us first recall that the $\Gamma $-functional is
the Legendre transform of $W(J,L)$ with respect to $J$, 
\[
\Gamma (\Phi ,L)=-W(J,L)+\int J\Phi ,\qquad \Phi =\frac{\delta W}{\delta J}. 
\]
In this operation, the sources $L$ are just spectators, so we have $\delta
\Gamma /\delta L_{I}=-\delta W/\delta L_{I}$.

Now, assuming that the functional $W$ is the improved one, we define the
master functional $\Omega (\Phi ,N)$ as the Legendre transform of $W(J,L)$
with respect to both $J$ and $L$, namely 
\begin{equation}
\Omega (\Phi ,N)=-W(J,L)+\int J\Phi +\int L_{I}N^{I},  \label{definemast}
\end{equation}
where 
\begin{equation}
\Phi =\frac{\delta W}{\delta J},\qquad N^{I}=\frac{\delta W}{\delta L_{I}}.
\label{relaf0}
\end{equation}
Clearly, $\Omega $ is also the Legendre transform of minus $\Gamma (\Phi ,L)$
with respect to $L$: 
\begin{equation}
\Omega (\Phi ,N)=\Gamma (\Phi ,L)+\int L_{I}N^{I},  \label{biu}
\end{equation}
where 
\begin{equation}
N^{I}=-\frac{\delta \Gamma }{\delta L_{I}}.  \label{biu2}
\end{equation}
We have the inverse formulas 
\begin{equation}
J=\frac{\delta \Omega }{\delta \Phi },\qquad L_{I}=\frac{\delta \Omega }{%
\delta N^{I}}.  \label{clearly}
\end{equation}

Let us show that $\Omega $ is indeed well-defined and collects one-particle
irreducible diagrams. To achieve this goal, it is convenient to view $\Omega 
$ as the Legendre transform (\ref{biu}) of minus $\Gamma $ with respect to $%
L $. We can use (\ref{biu2}) to expand $N(\Phi ,L)$ in powers of $L$. The
coefficients of this expansion are the (renormalized) connected,
one-particle irreducible correlation functions $\langle \mathcal{O}_{\mathrm{%
R}}^{I_{1}}\cdots \mathcal{O}_{\mathrm{R}}^{I_{n}}\rangle _{\text{1PI},L=0}$
(plus counterterms taking care of coinciding points), containing single or
multiple insertions of renormalized composite operators $\mathcal{O}_{%
\mathrm{R}}^{I}$. Using (\ref{sl}) we get 
\begin{equation}
N^{I}=(A^{-1})^{IJ}L_{J}+\langle \mathcal{O}_{\mathrm{R}}^{I}\rangle +\int 
\hat{\tau}_{vJ}\frac{\delta \mathcal{N}^{v}(L)}{\delta L_{I}}\langle 
\mathcal{O}_{\mathrm{R}}^{J}\rangle ,  \label{relaf01}
\end{equation}
whence 
\begin{equation}
\tilde{N}^{I}\equiv N^{I}-\langle \mathcal{O}_{\mathrm{R}}^{I}\rangle _{%
\text{1PI},L=0}=(A^{-1})^{IJ}L_{J}+\int \langle \mathcal{O}_{\mathrm{R}}^{I}%
\hspace{0.01in}\mathcal{O}_{\mathrm{R}}^{J}\rangle _{\text{1PI}%
,L=0}L_{J}+\int \hat{\tau}_{vJ}\frac{\delta \mathcal{N}^{v}(L)}{\delta L_{I}}%
\langle \mathcal{O}_{\mathrm{R}}^{J}\rangle _{\text{1PI},L=0}+\mathcal{O}%
(L^{2}).  \label{relaf1}
\end{equation}
Formula (\ref{assigne1}) tells us that the quantities $\tilde{N}^{I}$ are $%
\mathcal{O}(\delta ^{-n_{I}})$. The improvement term (\ref{improfin}) is
responsible for the contribution $A^{-1}L$ appearing on the right-hand side
of (\ref{relaf1}), which is crucial for the invertibility of (\ref{relaf1}).
Expanding in orders of $\delta $ we can invert (\ref{relaf1}) and find 
\begin{equation}
L_{I}(\Phi ,N)=A_{IJ}\tilde{N}^{J}-A_{IJ}A_{KH}\int \langle \mathcal{O}_{%
\mathrm{R}}^{J}\hspace{0.01in}\mathcal{O}_{\mathrm{R}}^{K}\rangle _{\text{1PI%
},L=0}\tilde{N}^{H}-\int \hat{\tau}_{vK}\frac{\delta \mathcal{N}^{v}(A\tilde{%
N})}{\delta \tilde{N}^{I}}\langle \mathcal{O}_{\mathrm{R}}^{K}\rangle _{%
\text{1PI},L=0}+\mathcal{O}(A^{3})\mathcal{O}(\tilde{N}).  \label{defineL}
\end{equation}

Now we are ready to prove that the master functional $\Omega $ just contains
one-particle irreducible diagrams glued together as shown in the pictures 
\begin{equation}
\includegraphics[width=3.5truein,height=1truein]{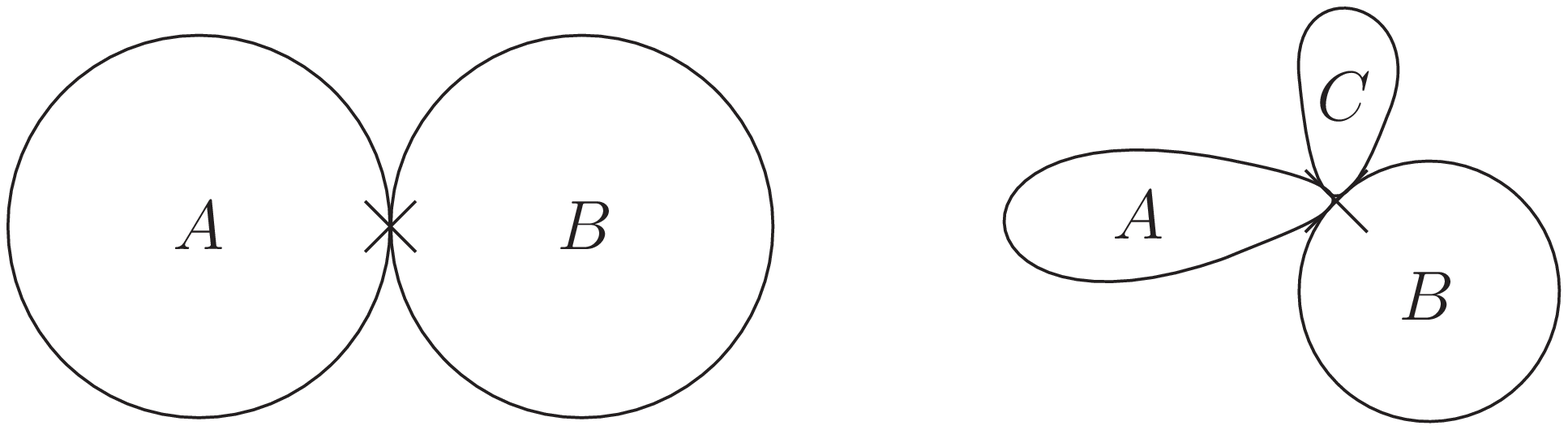}
\label{pitture}
\end{equation}
Here $A$, $B$ and $C$ can be any correlation functions $\langle \mathcal{O}_{%
\mathrm{R}}^{I_{1}}\cdots \mathcal{O}_{\mathrm{R}}^{I_{n}}\rangle _{\text{1PI%
},L=0}$, while the symbol $\times $ denotes that two or more composite-field
insertions are ``locally connected'' using vertices provided by $\mathcal{N}%
^{v}(L)$ and the ``identity propagators'' provided by (\ref{improfin}).

Consider first $\Omega =\Omega (\Phi ,N(\Phi ,L))$ as a functional of $\Phi $
and $L$, as given by the right-hand side of (\ref{biu}). This expression is
a generating functional of one-particle irreducible diagrams in the same way
as $\Gamma $ is. Indeed, because of (\ref{biu2}), the right-hand side of (%
\ref{biu}) collects the same correlation functions that are contained inside 
$\Gamma $, however multiplied by different coefficients.

Now we express the sources $L$ as functions of $\Phi $ and $N$. Using (\ref
{defineL}) we see that we get precisely the objects depicted in the pictures
(\ref{pitture}). In momentum space we have just products of correlation
functions $\langle \mathcal{O}_{\mathrm{R}}^{I_{1}}\cdots \mathcal{O}_{%
\mathrm{R}}^{I_{n}}\rangle _{\text{1PI},L=0}$ and polynomials. This argument
proves that the master functional obeys the theorem of locality of
counterterms. For the moment we are satisfied with this result. Later, in
section 7, we develop a ``proper formalism'' that allows us to study $\Omega 
$ using diagrammatic rules analogous to the ones we normally use for $\Gamma 
$, in particular calculate the renormalization of $\Omega $ working directly
on $\Omega $ without using the definitions (\ref{definemast}) and (\ref{biu}%
) based on $W$ and $\Gamma $.

Let us discuss how $\Omega $ depends on $A$ and $\tilde{N}$. Because of (\ref
{defineL}), $L$ contains only powers $A^{m}\tilde{N}^{n}$ with $m\geqslant n$%
. More precisely, $L_{I}=A_{IJ}\tilde{N}^{J}$ plus a sum of powers $A^{m}%
\tilde{N}^{n}$ with $m\geqslant n+1$. Instead, due to the improvement term (%
\ref{improfin}) the $\tilde{N}$-dependence inside $\Omega $ has the form of
monomials $A^{m}\tilde{N}^{n}$ with $m\geqslant n-1$. More precisely, we can
write 
\begin{equation}
\Omega (\Phi ,N)=\Gamma (\Phi )+T_{\Omega }(\tilde{N})+\Delta _{2}\Omega
(\Phi ,\tilde{N}).  \label{struttu}
\end{equation}
where 
\begin{equation}
T_{\Omega }(\tilde{N})=\frac{1}{2}\int \tilde{N}^{I}A_{IJ}\tilde{N}^{J}
\label{improomo}
\end{equation}
is the $\Omega $-improvement term and $\Delta _{2}\Omega $ is a sum of
monomials of the form 
\[
(A_{I_{1}J_{1}}\tilde{N}^{J_{1}})\cdots (A_{I_{n}J_{n}}\tilde{N}^{J_{n}})%
\hspace{0.01in}X^{I_{1}\cdots I_{n}}(\Phi ) 
\]
with $n\geqslant 2$, where the $X$s are power series in $A$, and can contain
derivatives acting on $\Phi $ and on the $\tilde{N}$s. Of course, $\Delta
_{2}\Omega $ is of higher order in $\delta $ than $T_{\Omega }$. Note that
the term linear in $\tilde{N}$ is missing in (\ref{struttu}). Actually, we
introduced $\tilde{N}$ precisely to make this happen.

The functional $\Gamma (\Phi )$ is the minimum of $\Omega $ with respect to
the $N^{I}$s. Indeed, the conditions 
\begin{equation}
\frac{\delta \Omega }{\delta N^{I}}=0  \label{condamas}
\end{equation}
are nothing but $L_{I}=0$. The solutions of (\ref{condamas}) determine $%
N^{I} $ as functions of $\Phi $. Formula (\ref{struttu}) immediately gives $%
\tilde{N}^{I}=0$, or $N^{I}=\langle \mathcal{O}_{\mathrm{R}}^{I}\rangle
_{L=0}$, so finally 
\[
\Gamma (\Phi )=\Omega (\Phi ,\langle \mathcal{O}_{\mathrm{R}}^{I}\rangle
_{L=0}). 
\]
Another way to derive $\Gamma (\Phi )$ from $\Omega (\Phi ,N)$ is to take
the limit $A\rightarrow 0$, which is regular in $\Omega $ and is equivalent
to set $L_{I}=0$: 
\[
\Gamma (\Phi )=\lim_{A\rightarrow 0}\Omega (\Phi ,N). 
\]

So far we have been working with renormalized quantities, but every argument
can be applied to bare quantities with obvious modifications.

\paragraph{Example\newline
}

To give an explicit example, we consider a free massless scalar field and
the composite field $\varphi ^{2}/2$ coupled to the source $L_{2}$. We want
to work out the master functional to the order $\tilde{N}^{3}$. Let $L_{0}$
and $L_{1}$ denote the sources coupled with the identity operator and the
elementary field, as usual. We choose $A=$diag$(a_{0}\mu ^{-\varepsilon
},a_{1},a_{2}\mu ^{\varepsilon })$, where the factors $\mu ^{\varepsilon }$
are introduced to make the dimensions of $a_{0}$, $a_{1}$ and $a_{2}$
integer. The functional $W$ is easy to calculate (check for example section
12 of \cite{fieldcov}). We find 
\begin{eqnarray*}
W(J,L) &=&\frac{1}{2}\int \left\{ (J+L_{1})\frac{1}{-\Box -L_{2}}%
(J+L_{1})+\mu ^{-\varepsilon }\left( \frac{1}{a_{2}}+\delta _{a}\right)
L_{2}^{2}\right\} \\
&&-\frac{1}{2}\mathrm{tr}\ln (-\Box -L_{2})+\int L_{0}+\mu ^{\varepsilon
}\int \frac{L_{0}^{2}}{2a_{0}}+\int \frac{L_{1}^{2}}{2a_{1}},
\end{eqnarray*}
where $\delta _{a}=-(16\pi ^{2}\varepsilon )^{-1}$. Then 
\[
\Phi =\int \frac{1}{-\Box -L_{2}}(J+L_{1}),\qquad N_{0}=1+\mu ^{\varepsilon }%
\frac{L_{0}}{a_{0}},\qquad N_{1}=\Phi +\frac{L_{1}}{a_{1}}, 
\]
and, in momentum space, 
\begin{equation}
\tilde{N}_{2}(k)=N_{2}(k)-\frac{\Phi ^{2}(k)}{2}=\frac{\mu ^{-\varepsilon }}{%
a_{2}(k)}L_{2}(k)+\frac{1}{2}\int \mathrm{d}k^{\prime }G_{3}(k,k^{\prime
})L_{2}(k^{\prime })L_{2}(k-k^{\prime })+\mathcal{O}(L_{2}^{3}),
\label{bifu}
\end{equation}
where $G_{3}=\langle \varphi ^{2}\varphi ^{2}\varphi ^{2}\rangle /8$, $%
\mathrm{d}k^{\prime }$ stands for $\mathrm{d}^{D}k^{\prime }/(2\pi )^{D}$
and we have defined the running coupling 
\[
\frac{1}{a_{2}(k)}=\frac{1}{a_{2}}-\frac{1}{32\pi ^{2}}\ln \frac{k^{2}}{\mu
^{2}}. 
\]
Inverting the $N$-$L$ relations we find $L_{0}=\mu ^{-\varepsilon
}a_{0}(N_{0}-1)$, $L_{1}=a_{1}(N_{1}-\Phi )$ and 
\[
L_{2}(k)=\mu ^{\varepsilon }a_{2}(k)\tilde{N}_{2}(k)-\frac{1}{2}\mu
^{3\varepsilon }a_{2}(k)\int \mathrm{d}k^{\prime }G_{3}(k,k^{\prime
})a_{2}(k^{\prime })\tilde{N}_{2}(k^{\prime })a_{2}(k-k^{\prime })\tilde{N}%
_{2}(k-k^{\prime })+\mathcal{O}(a_{2}^{4})\mathcal{O}(\tilde{N}_{2}^{3}). 
\]
The functional $\Omega $ is 
\begin{eqnarray}
\Omega (\Phi ,N) &=&\int \frac{1}{2}(\partial _{\mu }\Phi )^{2}+\frac{%
a_{0}\mu ^{-\varepsilon }}{2}\int (N_{0}-1)^{2}+\frac{a_{1}}{2}\int
(N_{1}-\Phi )^{2}+\frac{\mu ^{\varepsilon }}{2}\int \mathrm{d}k\hspace{0.01in%
}\tilde{N}_{2(-k)}a_{2}(k)\tilde{N}_{2k}  \nonumber \\
&&\!\!\!\!\!\!\!\!\!\!\!\!{-\frac{\mu ^{3\varepsilon }}{6}\int \mathrm{d}k%
\mathrm{d}k^{\prime }G_{3}(k,k^{\prime })a_{2}(k)\tilde{N}%
_{2k}a_{2}(k^{\prime })\tilde{N}_{2k^{\prime }}a_{2}(k-k^{\prime })\tilde{N}%
_{2(k-k^{\prime })}+\mathcal{O}(a_{2}^{4})\mathcal{O}(\tilde{N}_{2}^{4}).}
\label{omica}
\end{eqnarray}
Clearly, the limit $a\rightarrow 0$ gives back the $\Gamma $-functional of
the free-field theory.

\section{Changes of field variables in the master functional}

\setcounter{equation}{0}

In this section we study the changes of field variables in the master
functional, using the redundant linear approach. Again, we work with
renormalized quantities, since the analysis of bare changes of variables is
practically identical.

In ref. \cite{fieldcov} it was explained that a change of field variables is
made of the source-redefinitions (\ref{bibip}), or 
\begin{equation}
L_{I}^{\prime }=(L_{J}-b_{J}J)(z^{-1})_{I}^{J},\qquad J^{\prime }=J,
\label{baremu}
\end{equation}
in the $Z$- and $W$-functionals, and that such functionals behave as
scalars. To make $L_{J}$ and $b_{J}J$ of the same $\delta $-order in (\ref
{baremu}), we must assume $b_{I}=\mathcal{O}(\delta ^{n_{I}-1})$. It is very
simple to work out how (\ref{baremu}) reflects in the $\Omega $-functional.
From (\ref{mapzw}) we have $W^{\prime }(J^{\prime },L^{\prime
})=W(J,L_{J}^{\prime }z_{I}^{J}+b_{I}J)$, so definitions (\ref{relaf0}) give 
\begin{equation}
\Phi ^{\prime }=\left. \frac{\delta W^{\prime }}{\delta J^{\prime }}\right|
_{L^{\prime }}=\Phi +b_{I}N^{I},\qquad N^{I\hspace{0.01in}\prime }=\left. 
\frac{\delta W^{\prime }}{\delta L_{I}^{\prime }}\right| _{J^{\prime
}}=z_{J}^{I}N^{J}.  \label{gig}
\end{equation}
Then (\ref{definemast}) gives 
\[
\Omega ^{\prime }(\Phi ^{\prime },N^{\prime })=-W^{\prime }(J^{\prime
},L^{\prime })+\int J^{\prime }\Phi ^{\prime }+\int L_{I}^{\prime }N^{I%
\hspace{0.01in}\prime }=-W(J,L)+\int J\Phi +\int L_{I}N^{I}=\Omega (\Phi
,N), 
\]
which shows that the master functional, differently from $\Gamma $, does
transform as expected. Note that the transformations (\ref{gig}) are linear
in $\Phi $ and $N$.

In \cite{fieldcov} it was also shown that redefinitions (\ref{baremu}) are
associated with a change of variables $\varphi ^{\prime }=\varphi ^{\prime
}(\varphi ,\lambda ,J,L)$ in the functional integral and a number of
parameter-redefinitions, e.g. $b^{\prime }=b^{\prime }(b,\tau ,\lambda ,\mu
) $, $\tau ^{\prime }=\tau ^{\prime }(b,\tau ,\lambda ,\mu )$. Of course
such reparametrizations must be finite, because they act on a convergent
functional.

In this paper we have split the set of parameters $\tau $ into $A^{-1}$ plus
the rest, and the rest was still called $\tau $. The two subsets play a
different role, because the improvement term is dominant with respect to the
other terms belonging to the source sector. Because of this, we have also
modified the $\delta $-assignments into (\ref{assigne1}). Thus, the
parameter-redefinitions associated with (\ref{baremu}) now read $A^{\prime 
\hspace{0.01in}-1}=A^{\prime \hspace{0.01in}-1}(b,A^{-1},\tau ,\lambda ,\mu
) $, $\tau ^{\prime }=\tau ^{\prime }(b,A^{-1},\tau ,\lambda ,\mu )$, etc.,
and must be determined carefully, because the change of variables makes $A$%
-denominators spread out everywhere. We must determine $A^{\prime }$ and $%
\tau ^{\prime }$ such that all $A^{\prime }$-denominators cancel out inside $%
\Omega ^{\prime }$. Then the limit $A^{\prime }\rightarrow 0$ of $\Omega
^{\prime }$ gives $\Gamma ^{\prime }$.

Separating the improvement term $T$ from the rest let us write 
\begin{equation}
W(J,L)=\tilde{W}(J,L)+T(L),  \label{gyk}
\end{equation}
where $\tilde{W}$ does not depend on $A$. When we make the substitutions (%
\ref{baremu}) we obtain 
\begin{equation}
W^{\prime }(J^{\prime },L^{\prime })=W(J,L)=\tilde{W}(J^{\prime },L^{\prime
}z+bJ)+\frac{1}{2}\int (L^{\prime }z+bJ^{\prime
})_{I}(A^{-1})^{IJ}(L^{\prime }z+bJ^{\prime })_{J}.  \label{gyk2}
\end{equation}
The last term of this formula contains powers $b^{m}A^{-n}$, with $%
m\geqslant n$. Working out $\Omega ^{\prime }$ from its definition (\ref
{definemast}) these powers spread out everywhere inside the transformed
master functional. From the point of view of the expansion in powers of $%
\delta $, negative $A$-powers are not a problem, since in any case the
orders of $\delta $ organize correctly. However, we want to be able to treat
the change of variables perturbatively, while $A$ is also treated
perturbatively. For example, it is sufficient to imagine that each $b_{I}$
carries an extra small parameter $\zeta $ besides the order of $\mathcal{O}%
(\delta ^{n_{I}-1})$ assigned to it, and expand in $\zeta $ before expanding
in $\delta $.

We can also view the problem of negative $A$-powers in the field
transformations (\ref{gig}). Those transformations do leave $\Omega ^{\prime
}$ regular for $A\rightarrow 0$, but they do not preserve the structure (\ref
{struttu}). In particular, they generate terms linear in $\tilde{N}$, which
are absent in (\ref{struttu}). To recover the primed version of (\ref
{struttu}) we must redefine $\tilde{N}$. However, it is easy to see that
when we do this, powers $b^{m}A^{-n}$, with $m\geqslant n$, propagate from
the improvement term to $\Gamma ^{\prime }(\Phi ^{\prime })$, $T_{\Omega
}^{\prime }(\tilde{N}^{\prime })$ and $\Delta _{2}\Omega ^{\prime }(\Phi
^{\prime },\tilde{N}^{\prime })$. To completely determine $\Omega ^{\prime
}(\Phi ^{\prime },N^{\prime })$ we must determine the parameters $b^{\prime
} $, $A^{\prime }$, $\tau ^{\prime }$ and the constants $z$ as functions of $%
A$, $b$, and $\tau $, so that they absorb away all negative $A$-powers and
turn the structure of $\Omega ^{\prime }(\Phi ^{\prime },N^{\prime })$ into
the primed version of (\ref{struttu}), where $\tilde{N}^{\prime }$ is worked
out solving $\delta \Omega ^{\prime }/\delta N^{\prime }=0$. Note that the
matrix $z$ is not uniquely determined, because after eliminating the
negative $A$-powers we can always make a further change of composite-field
basis.

Finally, we can also view this problem inside the functional integral, going
through section 10 of ref. \cite{fieldcov}. If the starting functional
integral is written in the conventional form, as we assume, the redefinition
(\ref{baremu}) turns it into some unconventional form. We can recover the
conventional form applying the theorem proved in section 9 of ref. \cite
{fieldcov}, but then it is easy to see that powers $b^{m}A^{-n}$, with $%
m\geqslant n$, propagate inside the change of field variables $\varphi
^{\prime }=\varphi ^{\prime }(\varphi ,\lambda ,J,L)$, as well as in $z$, $%
A^{\prime \hspace{0.01in}}$ and $\tau ^{\prime }$.

Now we give a step-by-step procedure to work out the reparametrization that
must accompany the change of field variables (\ref{gig}) to reabsorb the
negative $A$-powers. We work directly on the master functional, bypassing $Z$
and $W$. At the end of this section we illustrate the procedure with an
explicit example.

1) First we make the substitutions $\Phi =\Phi ^{\prime }-bz^{-1}N^{\prime }$%
, $N=z^{-1}N^{\prime }$ inside $\Omega (\Phi ,N)$. They do give the
transformed functional $\Omega ^{\prime }$, but this $\Omega ^{\prime }$ is
still written in the old parametrization. Next, we solve the conditions $%
\delta \Omega ^{\prime }/\delta N^{\prime \hspace{0.01in}I}=0$ and insert
the solutions $N^{\prime }(\Phi ^{\prime })$ back into $\Omega ^{\prime }$.
This operation gives $\Gamma ^{\prime }(\Phi ^{\prime })$, still written in
the old parametrization. We know, from the analysis of section \ref
{unexpected}, that there exists a non-local change of field variables $\Phi
^{\prime }(\Phi )$ such that $\Gamma ^{\prime }(\Phi ^{\prime })=\Gamma
(\Phi )$. The classical limits of $\Gamma (\Phi )$ and $\Gamma ^{\prime
}(\Phi ^{\prime })$ are the classical actions $S_{c}(\varphi )$ and $%
S_{c}^{\prime }(\varphi ^{\prime })$, before and after the change of
variables. They are related by the classical limit $\varphi ^{\prime
}(\varphi )$ of $\Phi ^{\prime }(\Phi )$. Inverting this relation and
writing it as 
\[
\varphi (\varphi ^{\prime })=\varphi ^{\prime }-b_{I}^{\prime }\mathcal{O}%
_{c}^{\prime \hspace{0.01in}I}(\varphi ^{\prime }), 
\]
we determine the constants $b_{I}^{\prime }$. They make $S_{c}^{\prime
}(\varphi ^{\prime })$ free of $A$-denominators, because $S_{c}(\varphi )$
is independent of $A$.

2) At this point, we consider again the solutions $N^{\prime }(\Phi ^{\prime
})$ of $\delta \Omega ^{\prime }/\delta N^{\prime \hspace{0.01in}I}=0$.
These are the average values $\langle \mathcal{O}_{\mathrm{R}}^{I\hspace{%
0.01in}\prime }\rangle ^{\prime }$ in the new variable frame, at $L^{\prime
}=0$, and must also be regular. We determine the constants $z$ canceling the
negative $A$-powers of the $\langle \mathcal{O}_{\mathrm{R}}^{I\hspace{0.01in%
}\prime }\rangle ^{\prime }$-classical limits.

3) Finally, we are ready to consider $\Omega ^{\prime }(\Phi ^{\prime
},N^{\prime })$. The new parameters $A^{\prime }$ and $\tau ^{\prime }$ are
determined matching its structure with the primed version of (\ref{struttu}%
), again in the classical limit. Once we express $z$ and $A$, $b$, and $\tau 
$ as functions of $A^{\prime }$, $b^{\prime }$, and $\tau ^{\prime }$,
everywhere, we obtain the correctly parametrized $\Omega ^{\prime }$.

Observe that at each step we determine the desired reparametrizations
working with classical limits. Indeed, the reparametrization is fully
determined by those limits, in the same way as the entire functional $\Omega 
$ is fully determined by the classical action, by means of Feynman rules and
Feynman diagrams (see section 7). When the classical limits are matched,
radiative corrections automatically turn out to be right. Moreover, they are
consistent with the perturbative expansion in $\delta $.

We could also find the desired reparametrizations working with the
renormalized actions $S_{L}$ and $S_{L}^{\prime }$, instead of working with $%
\Omega $. However, it would not make much difference: the divergent parts
cannot enter the reparametrizations, which are finite, and once we drop them
we end up again matching the classical limits.

Summarizing, a change of variables in the master functional is the linear
redefinition 
\begin{equation}
\Phi ^{\prime }=\Phi +b_{I}N^{I},\qquad N^{I\hspace{0.01in}\prime
}=z_{J}^{I}N^{J},  \label{baremastr}
\end{equation}
under which $\Omega $ behaves as a scalar, $\Omega ^{\prime }(\Phi ^{\prime
},N^{\prime })=\Omega (\Phi ,N)$. To find the correct structure of $\Omega
^{\prime }$ we must accompany (\ref{baremastr}) with a set of
reparametrizations that can be worked out with the procedure outlined above.

Now we illustrate the main issues with the help of an example.

\paragraph{Example\newline
}

We consider again the free theory of a massless scalar field, with the
composite field $\varphi ^{2}/2$ coupled to the source $L_{2}$. We want to
study the change of variables $L=L^{\prime }z+bJ$ to the order $b^{2}$ in
the functionals $\Omega $ and $\Gamma $ and check the results computing the
associated Feynman diagrams. We treat $\tilde{N}$ as an $\mathcal{O}(b)$%
-object and truncate the $\Omega $-functional to the first line of (\ref
{omica}). In this approximation the field transformation and the functional $%
\Gamma ^{\prime }$ can be calculated up to $\mathcal{O}(b^{2})$, while $%
\tilde{N}^{\prime }$ can be worked out up to $\mathcal{O}(b)$.

The change of variables reads 
\begin{equation}
\Phi ^{\prime }=\Phi +b_{2}N_{2}+b_{1}N_{1},\qquad N_{2}^{\prime
}=z_{22}N_{2}+z_{21}N_{1},\qquad N_{1}^{\prime
}=z_{11}N_{1}+z_{12}N_{2},\qquad N_{0}^{\prime }=N_{0},  \label{substi}
\end{equation}
and the transformed $\Omega $-functional $\Omega ^{\prime }$ is $\Omega
(\Phi ,N)$ once (\ref{substi}) are implemented. To find the correct
reparametrizations, we first solve the conditions $\delta \Omega ^{\prime
}/\delta N_{2}^{\prime }=\delta \Omega ^{\prime }/\delta N_{1}^{\prime
}=\delta \Omega ^{\prime }/\delta N_{0}^{\prime }=0$. Inserting the
solutions $\tilde{N}^{\prime }(\Phi ^{\prime })$ back inside $\Omega
^{\prime }$ we get $\Gamma ^{\prime }(\Phi ^{\prime })$. Then it is
relatively easy to check that 
\begin{equation}
\Gamma ^{\prime }(\Phi ^{\prime })=\frac{1}{2}\int \mathrm{d}^{D}x\
(\partial _{\mu }\Phi (\Phi ^{\prime }))^{2},  \label{bisu}
\end{equation}
where 
\begin{equation}
\Phi (\Phi ^{\prime })=\Phi ^{\prime }(1-b_{1}^{\prime })-\frac{%
b_{2}^{\prime }}{2}\Phi ^{\prime \hspace{0.01in}2}+\frac{b_{2}^{\prime 
\hspace{0.01in}2}}{2}\Phi ^{\prime \hspace{0.01in}3}-\frac{b_{2}^{\prime 
\hspace{0.01in}2}\mu ^{-\varepsilon }}{64\pi ^{2}}\left( \ln \frac{-\Box }{%
\mu ^{2}}\right) (\Box \Phi ^{\prime })+\mathcal{O}(b^{3}).  \label{biso}
\end{equation}
The relations between $b$ and $b^{\prime }$ are 
\[
b_{1}=b_{1}^{\prime }+b_{1}^{\prime \hspace{0.01in}2}+\frac{1}{2}\left( 
\frac{b_{1}^{\prime \hspace{0.01in}2}}{a_{1}}+\frac{b_{2}^{\prime \hspace{%
0.01in}2}\mu ^{-\varepsilon }}{a_{2}}\right) \Box +\mathcal{O}(b^{3}),\qquad
b_{2}=b_{2}^{\prime }+3b_{1}^{\prime }b_{2}^{\prime }+\mathcal{O}(b^{3}). 
\]
Boxes appear inside our ``constants'' because we work in an approach where
descendants, such as $\Box \varphi $, $\Box \varphi ^{2}$, $\Box ^{2}\varphi 
$, etc., are not viewed as independent composite fields, but treated
altogether with their primaries. This amounts to promote the constants to
polynomials in derivatives. Note that formula (\ref{biso}) contains also the
cubic power of the field. Since we have not introduced an independent source
for the composite field $\varphi ^{3}$, the coefficient of $\Phi ^{\prime 
\hspace{0.01in}3}$ in (\ref{biso}) is not independent, but a function of $%
b_{1}^{\prime }$ and $b_{2}^{\prime }$.

Clearly, the classical limits of (\ref{bisu}) and (\ref{biso}) are local. It
is easy to check by explicit computation that the radiative corrections of (%
\ref{bisu}) are determined by the classical limit of (\ref{bisu}) in the
usual way. There is just one one-loop diagram to compute, the scalar
self-energy made with two vertices $(b_{2}^{\prime }/2)\varphi ^{\prime 
\hspace{0.01in}2}\Box \varphi ^{\prime }$.

Observe that (\ref{biso}) is also the appropriate non-local variable change
of the $\Gamma $-functional, that is to say (\ref{nonlocchv}) at $L^{\prime
}=0$ (upon converting the notation of that formula to the one used here).

We have worked out the reparametrizations $b^{\prime }(b)$ that make all $a$%
-denominators disappear from $\Gamma ^{\prime }(\Phi ^{\prime })$. The next
task is to find the values of $z_{ij}$ that reabsorb the $a$-denominators
contained in the averages $\langle \mathcal{O}_{\mathrm{R}}^{I\hspace{0.01in}%
\prime }\rangle _{L^{\prime }=0}^{\prime }$. This is straightforward, since
we already have such averages from the solutions of $\delta \Omega ^{\prime
}/\delta N^{I\hspace{0.01in}\prime }=0$. Proceeding order-by-order in $b$ we
find 
\begin{eqnarray}
\mathcal{O}^{1\hspace{0.01in}\prime } &=&\varphi ^{\prime },\qquad \mathcal{O%
}^{2\hspace{0.01in}\prime }=\frac{\varphi ^{\prime \hspace{0.01in}2}}{2}-%
\frac{b_{2}^{\prime }}{2}\mu ^{\varepsilon /2}\varphi ^{\prime \hspace{0.01in%
}3}+\mathcal{O}(b^{2}),  \label{nec} \\
\langle \mathcal{O}_{\mathrm{R}}^{2\hspace{0.01in}\prime }\rangle
_{L^{\prime }=0}^{\prime } &=&\frac{\Phi ^{\prime \hspace{0.01in}2}}{2}-%
\frac{b_{2}^{\prime }}{2}\Phi ^{\prime \hspace{0.01in}3}+\frac{b_{2}^{\prime
}\mu ^{-\varepsilon }}{32\pi ^{2}}\left( \ln \frac{-\Box }{\mu ^{2}}\right)
(\Box \Phi ^{\prime })+\mathcal{O}(b^{2}),  \label{neco}
\end{eqnarray}
together with 
\[
z_{11}=1+b_{1}^{\prime }+\frac{b_{1}^{\prime }}{a_{1}}\Box +\mathcal{O}%
(b^{2}),\quad z_{12}=b_{2}^{\prime }+\mathcal{O}(b^{2}),\quad z_{21}=\frac{%
b_{2}^{\prime }}{a_{2}}\mu ^{-\varepsilon }\Box +\mathcal{O}(b^{2}),\quad
z_{22}=1+2b_{1}^{\prime }+\mathcal{O}(b^{2}). 
\]
Again, it is easy to check by explicit computation that the radiative
corrections contained in (\ref{neco}) are those predicted by the new
classical action and the new composite fields (\ref{nec}).

The final task is to find the reparametrizations $A^{\prime }$ and $\tau
^{\prime }$ that make $\Omega ^{\prime }$ have the correct dependence on $%
A^{\prime }$ and $\tilde{N}^{\prime }$, which is encoded in the primed
version of formula (\ref{struttu}). In our approximation we have to stop at
the terms that are quadratic in $\tilde{N}^{\prime }$ and $\mathcal{O}%
(b^{0}) $. We find 
\[
\Omega ^{\prime }(\Phi ^{\prime },N^{\prime })=\Gamma ^{\prime }(\Phi
^{\prime })+\frac{1}{2}\int \tilde{N}^{\prime \hspace{0.01in}%
I}A_{IJ}^{\prime }\tilde{N}^{\prime \hspace{0.01in}J}+\mathcal{O}(b)\mathcal{%
O}(\tilde{N}^{\prime \hspace{0.01in}2})+\mathcal{O}(\tilde{N}^{\prime 
\hspace{0.01in}3}), 
\]
where $A_{IJ}^{\prime }=$diag$(a_{0}\mu ^{-\varepsilon },a_{1},a_{2}(k)\mu
^{\varepsilon })+\mathcal{O}(b)$.

As expected, the new parametrization, obtained matching only tree-level
contributions, makes all terms regular inside $\Omega ^{\prime }$, including
radiative corrections.

\section{Restrictions}

\setcounter{equation}{0}

The sources $L_{I}$ and their ``Legendre-partners'' $N^{I}$ are useful tools
to study composite fields and field redefinitions, but at some point we may
want to get rid of them choosing suitable restrictions and define some sort
of ``quantum action'' $\Omega (\Phi )$ depending only on the fields $\Phi $.
In this section we consider some options of this kind. The $\Gamma $%
-functional can be viewed as one of them.

Choose a restriction $N^{I}=N^{I}(\Phi )$, where the functions $N^{I}(\Phi )$
are unspecified for the moment, and define $\Omega (\Phi )=\Omega (\Phi
,N(\Phi ))$. Because of (\ref{baremastr}) the transformed restriction is $%
N^{I}(\Phi )=(z^{-1})_{J}^{I}N^{J\hspace{0.01in}\prime }(\Phi ^{\prime })$.
The change of variables reads 
\[
\Phi ^{\prime }(\Phi )=\Phi +b_{I}N^{I}(\Phi ) 
\]
and the restricted master functional transforms correctly, 
\[
\Omega ^{\prime }(\Phi ^{\prime })=\Omega ^{\prime }(\Phi ^{\prime
},N^{\prime }(\Phi ^{\prime }))=\Omega (\Phi ,N(\Phi ))=\Omega (\Phi ). 
\]

A simple restriction is $N^{I\hspace{0.01in}\prime }=N^{I}=0$, however in
this case the field redefinition (\ref{baremastr}) is just the identity $%
\Phi ^{\prime }=\Phi $. The restriction $L_{I}=0$ or, equivalently, $%
N^{I}=\langle \mathcal{O}_{\mathrm{R}}^{I}\rangle _{L=0}=N^{I}(\Phi )$,
gives the functional $\Gamma (\Phi )$. In that case the change of variables
becomes 
\begin{equation}
\Phi ^{\prime }(\Phi )=\Phi +b_{I}\langle \mathcal{O}_{\mathrm{R}%
}^{I}\rangle _{L=0}  \label{expre}
\end{equation}
and we have 
\[
\Gamma (\Phi )=\Omega (\Phi )=\Omega (\Phi ,N(\Phi ))=\Omega ^{\prime }(\Phi
^{\prime },N^{\prime }(\Phi ^{\prime }))=\Omega ^{\prime }(\Phi ^{\prime }). 
\]
However, the last expression does not coincide with $\Gamma ^{\prime }(\Phi
^{\prime })$. Indeed, we know that, although the restricted master
functional does transform correctly, $\Gamma $ does not transform as
expected. We get the correct transformed $\Gamma $-functional $\Gamma
^{\prime }(\Phi ^{\prime })$ when the restriction reads $N^{I\hspace{0.01in}%
\prime }=\langle \mathcal{O}_{\mathrm{R}}^{I\hspace{0.01in}\prime }\rangle
_{L^{\prime }=0}$ in the new variables, or $L_{I}^{\prime }=0$, but (\ref
{bibip}) shows that $L_{I}=0$ cannot imply $L_{I}^{\prime }=0$. Applying the
change of variables we find instead that the transformed restriction reads $%
N^{I\hspace{0.01in}\prime }=z_{J}^{I}\langle \mathcal{O}_{\mathrm{R}%
}^{J}\rangle _{L=0}$.

To recover the correct transformed $\Gamma $-functional we must make an
additional step, similar to the one explained in section \ref{unexpected}.
Consider the difference 
\begin{eqnarray}
\tilde{N}^{I\hspace{0.01in}\prime } &=&z_{J}^{I}\langle \mathcal{O}_{\mathrm{%
R}}^{J}\rangle _{L=0}-\langle \mathcal{O}_{\mathrm{R}}^{I\hspace{0.01in}%
\prime }\rangle _{L^{\prime }=0}=z_{J}^{I}\left. \frac{\delta W}{\delta L_{I}%
}\right| _{L=0}-\left. \frac{\delta W^{\prime }}{\delta L_{I}^{\prime }}%
\right| _{L^{\prime }=0}  \nonumber \\
&=&\left. \frac{\delta W^{\prime }}{\delta L_{I}^{\prime }}\right|
_{L^{\prime }=-bz^{-1}J}-\left. \frac{\delta W^{\prime }}{\delta
L_{I}^{\prime }}\right| _{L^{\prime }=0}=\left. \frac{\delta \Gamma ^{\prime
}(\Phi ^{\prime },L^{\prime })}{\delta L_{I}^{\prime }}\right| _{L^{\prime
}=0}\left. -\frac{\delta \Gamma ^{\prime }(\Phi ^{\prime },L^{\prime })}{%
\delta L_{I}^{\prime }}\right| _{L^{\prime }=-bz^{-1}J}.  \label{buio}
\end{eqnarray}
Now, observe that at $L_{I}=0$, $J$ coincides with the field equations $%
\delta \Gamma (\Phi )/\delta \Phi $. Using (\ref{expre}) we can view the
right-hand side of (\ref{buio}) as a function of $\Phi $. Clearly, this
function is proportional to $J=\delta \Gamma (\Phi )/\delta \Phi $ and the
``coefficient'' of $J$ is a collection of one-particle irreducible diagrams.
Then, by the primed version of (\ref{struttu}) the difference 
\[
\Gamma (\Phi )-\Gamma ^{\prime }(\Phi ^{\prime })=\Omega ^{\prime }(\Phi
^{\prime })-\Gamma ^{\prime }(\Phi ^{\prime })=T_{\Omega }^{\prime }(\tilde{N%
}^{\prime }(\Phi ^{\prime }))+\Delta _{2}\Omega ^{\prime }(\Phi ^{\prime },%
\tilde{N}^{\prime }(\Phi ^{\prime })). 
\]
is quadratically proportional to $\tilde{N}^{\prime }$. By (\ref{buio}),
when expressed as a function of $\Phi $ it has the form 
\[
-\int \frac{\delta \Gamma (\Phi )}{\delta \Phi }\mathcal{M}(\Phi )\frac{%
\delta \Gamma (\Phi )}{\delta \Phi }, 
\]
namely it is quadratically proportional to the field equations $\delta
\Gamma (\Phi )/\delta \Phi $. Moreover, the ``coefficient of
proportionality'' $\mathcal{M}(\Phi )$ collects one-particle irreducible
diagrams and is local at the tree level. Then we can use the theorem
recalled in the appendix and absorb the difference $\Gamma -\Gamma ^{\prime
} $ inside a further change of variables $\tilde{\Phi}(\Phi )$, which is the
sum of a tree-level perturbative field redefinition plus one-particle
irreducible radiative corrections. Finally, we get $\Gamma ^{\prime }(\Phi
^{\prime })=\Gamma (\tilde{\Phi}(\Phi (\Phi ^{\prime })))=\tilde{\Gamma}(%
\tilde{\Phi})$, if we define $\Gamma \equiv \tilde{\Gamma}$. We find, as in
section 2, that the functionals $\Gamma $ and $\Gamma ^{\prime }$ are mapped
into each other, but the correct field transformation is not just (\ref
{expre}), rather $\tilde{\Phi}(\Phi (\Phi ^{\prime }))$. Clearly, this map
preserves the structure (\ref{gastru}).

Other restrictions $N^{I}(\Phi )$ may be useful for different purposes. For
example, if we choose $L_{I}=\ell _{I}=$constants, we turn the classical
action $S_{c}(\varphi )$ into $S_{c}(\varphi )-\sum_{I}\ell _{I}\mathcal{O}%
^{I}(\varphi )$. In this way we can study all actions, therefore all
theories with the same field content, at the same time.

\section{Proper formulation}

\setcounter{equation}{0}

In this section we show that with the help of a simple trick we can work
with the master functional in a more economic way. The action $S_{L}(\varphi
,L)$ appearing in the $Z$-integrand is not sufficiently similar to the
master functional $\Omega (\Phi ,N)$ and the classical action $%
S_{cL}(\varphi ,L)$ does not coincide with the classical limit of $\Omega $.
In particular, $S_{L}$ depends on ``mixed'' variables, since the sources $L$
are, strictly speaking, arguments of the functionals $Z$ and $W$, together
with $J$, not arguments of an action. We want an action $S_{N}(\varphi
,N_{S})$ that coincides with the master functional in the classical limit,
therefore it must depend on $\varphi $ and some new ``fields'' $N_{S}$, such
that $\Phi =\langle \varphi \rangle $ and $N=\langle N_{S}\rangle $. We call
this formulation the \textit{proper formulation} of the master functional.
Among the other things, it allows us to work directly on the master
functional from the very beginning, without passing from $Z$, $W$ or $\Gamma 
$. To study the renormalization of $\Omega $ it is sufficient to write the
Feynman rules of the \textit{proper action} $S_{N}(\varphi ,N_{S})$ and work
out their one-particle irreducible Feynman diagrams. Finally, in the proper
formulation the conventional form of the functional integral is manifestly
preserved during a general change of field variables.

To begin with, it is easy to see that the $Z$-functional (\ref{uy0}) can be
expressed in the form 
\begin{equation}
Z(J,L)=\int [\mathrm{d}\varphi \hspace{0.01in}\mathrm{d}N_{S}\hspace{0.01in}%
\mathrm{d}\tilde{L}]\exp \left( -S_{L}(\varphi ,\tilde{L})+\int J\varphi
+\int (L_{I}-\tilde{L}_{I})N_{S}^{I}\right) .  \label{uy1}
\end{equation}
Indeed, the $N_{S}$-integral gives a functional $\delta $-function $\delta
(L_{I}-\tilde{L}_{I})$ and the further $\tilde{L}$-integral gives back (\ref
{uy0}). Now, define the proper action $S_{N}(\varphi ,N_{S})$ from the
formula 
\begin{equation}
\exp \left( -S_{N}(\varphi ,N_{S})\right) \equiv \int [\mathrm{d}L]\exp
\left( -S_{L}(\varphi ,L)-\int L_{I}N_{S}^{I}\right) .  \label{uy2}
\end{equation}
Inserting (\ref{uy2}) with $L\rightarrow \tilde{L}$ in (\ref{uy1}) we can
express the $Z$- and $W$-functionals as 
\begin{equation}
Z(J,L)=\exp W(J,L)=\int [\mathrm{d}\varphi \hspace{0.01in}\mathrm{d}%
N_{S}]\exp \left( -S_{N}(\varphi ,N_{S})+\int J\varphi +\int
L_{I}N_{S}^{I}\right) .  \label{propz}
\end{equation}
Here each composite field is associated with an integrated variable $N_{S}$
and an external source $L$. Both $\varphi $ and $N_{S}$ are regarded as
elementary fields, called \textit{proper fields}.

The exponent $-S_{N}$ on the left-hand side of (\ref{uy2}) can be viewed as
the $W$-functional associated with the functional integral appearing on the
right-hand side of the same formula, where the fields $\varphi $ are treated
as external variables and the $L$-propagators are those provided by the
improvement term contained in $S_{L}$. The $L$-functional integral of (\ref
{uy2}) is a purely algebraic operation, because the $L$-propagators are
equal to the identity in momentum space. The loop diagrams are integrals of
the form 
\[
\int \frac{\mathrm{d}^{D}p}{(2\pi )^{D}}P(p), 
\]
where $P(p)$ is a polynomial, so they vanish using the dimensional
regularization. Thus the action $S_{N}$ receives only tree-level
contributions, therefore it is local.

We can work out $S_{N}$ explicitly using the saddle-point approximation,
which is actually exact in the case of the functional integral (\ref{uy2}).
Let $L_{I}=L_{I}^{*}(\varphi ,N_{S})$ denote the perturbative solutions of 
\[
N_{S}^{I}=-\frac{\delta S_{L}(\varphi ,L)}{\delta L_{I}}. 
\]
Then, writing $\tilde{L}=L-L^{*}$ and expanding the integrand of (\ref{uy2})
around $L^{*}(\varphi ,N_{S})$, the right-hand side of (\ref{uy2}) becomes 
\[
\int [\mathrm{d}\tilde{L}]\exp \left( -S_{L}(\varphi ,L^{*})-\int
L_{I}^{*}N_{S}^{I}+\mathcal{O}(\tilde{L}^{2})\right) =\exp \left(
-S_{L}(\varphi ,L^{*})-\int L_{I}^{*}N_{S}^{I}\right) . 
\]
The last expression is proved observing that the $\tilde{L}$-propagators are
equal to the identity, and the $\tilde{L}$-functional integral involves only
vertices that have at least two $\tilde{L}$-legs. So, it can receive
contributions only from loop diagrams, which however vanish. Finally, we get 
\begin{equation}
S_{N}(\varphi ,N_{S})=S_{L}(\varphi ,L^{*}(\varphi ,N_{S}))+\int
L_{I}^{*}(\varphi ,N_{S})N_{S}^{I}.  \label{saddlesol}
\end{equation}
In practice, $S_{N}$ coincides with the Legendre transform of $-S_{L}$ with
respect to $L$. In particular, we have the relation 
\[
\frac{\delta S_{N}}{\delta N^{I}}=L_{I}^{*}(\varphi ,N_{S}). 
\]

The inverse of formula (\ref{uy2}) reads 
\[
\exp \left( -S_{L}(\varphi ,L)\right) =\int [\mathrm{d}N_{S}]\exp \left(
-S_{N}(\varphi ,N_{S})+\int L_{I}N_{S}^{I}\right) . 
\]
The integral over $N_{S}$ can be calculated like the $L$-integral of (\ref
{uy2}), and receives only tree-level contributions because the $N_{S}$%
-propagators are also proportional to the identity. Alternatively, to go
from $S_{N}$ to $S_{L}$ we can use the inverse Legendre transform.

The proper formulation is convenient for several reasons, which we now
illustrate. The generating functionals $Z$ and $W$ associated with the
extended action $S_{L}$ (where the fields $\varphi $ are integrated and $L$
are external sources) can also be viewed as the generating functionals $Z$
and $W$ associated with the proper action $S_{N}$ (where both $\varphi $ and 
$N_{S}$ are integrated fields).

On the other hand, the master functional $\Omega $ can be viewed as the $%
\Gamma $-functional of the proper approach. Indeed, the master functional $%
\Omega $ is the Legendre transform of $W$ with respect to both $J$ and $L$.
In the proper approach this is precisely the $\Gamma $-functional, because
now the integrated fields are both $\varphi $ and $N_{S}$, while $J$ and $L$
are the sources coupled with them. Clearly, the classical limit of the
master functional $\Omega (\Phi ,N)$ coincides with the classical action $%
S_{cN}(\Phi ,N)$ of the proper approach, and $\Phi =\langle \varphi \rangle $%
, $N=\langle N_{S}\rangle $, as promised. Moreover, the master functional
has the structure (\ref{gastru}), which means that its radiative corrections
follows from its classical limit $S_{cN}$ according to the usual rules.

When no confusion can arise, we drop the subscript $S$ in $N_{S}$ and use
the symbol $N$ for the variables of $S_{N}$. Some other times we may denote
the $N$-variables of $\Omega $ with $N_{\Omega }$.

As a first example, we work out $S_{N}$ for the basic $S_{L}$-action 
\[
S_{0L}(\varphi ,L)=S(\varphi )-\int L_{I}\mathcal{O}_{\mathrm{R}%
}^{I}(\varphi )-\frac{1}{2}\int L_{I}(A^{-1})^{IJ}L_{J}. 
\]
The functional integral of (\ref{uy2}) is Gaussian and gives 
\[
S_{0N}(\varphi ,N)=S(\varphi )+\frac{1}{2}\int \tilde{N}^{I}A_{IJ}\tilde{N}%
^{J}, 
\]
where $\tilde{N}^{I}=N^{I}-\mathcal{O}_{\mathrm{R}}^{I}(\varphi )$. More
generally, we can work out $S_{N}$ either using (\ref{saddlesol}) or
expanding around $S_{0N}$. Decompose the complete action $S_{L}$ (\ref{sl})
as 
\begin{equation}
S_{L}(\varphi ,L)=S_{0L}(\varphi ,L)-\int \mathbf{\tau }_{vI}\mathcal{N}%
^{v}(L)\mathcal{O}_{\mathrm{R}}^{I}(\varphi ),  \label{uy3}
\end{equation}
where $S_{0L}$ is the part we expand around, while the terms $\mathbf{\tau }%
_{vJ}\mathcal{N}^{v}\mathcal{O}_{\mathrm{R}}^{J}$ are treated
perturbatively. The action $S_{N}$ is equal to $S_{0N}$ plus corrections
that we now describe. Inserting (\ref{uy3}) in (\ref{uy2}) and observing
that each $L$-insertion can be traded for minus the functional derivative $%
\delta /\delta N$ and moved outside of the functional integral, we can write
a formula that implicitly gives $S_{N}$. Precisely, 
\[
\exp \left( -S_{N}(\varphi ,N)\right) =\exp \left( \int \tau _{vJ}\mathcal{N}%
^{v}(-\delta /\delta N)\mathcal{O}_{\mathrm{R}}^{J}(\varphi )\right) \exp
\left( -S_{0N}(\varphi ,N)\right) . 
\]
Next, observe that $\delta S_{0N}/\delta N^{I}=A_{IJ}\tilde{N}^{J}$, so the
structure of $S_{N}$ is 
\begin{equation}
S_{N}(\varphi ,N)=S_{0N}(\varphi ,N)+\sum_{n\geqslant 0}(A_{I_{1}J_{1}}%
\tilde{N}^{J_{1}})\cdots (A_{I_{n}J_{n}}\tilde{N}^{J_{n}})\hspace{0.01in}%
\tilde{X}_{I}^{I_{1}\cdots I_{n}}\mathcal{O}_{\mathrm{R}}^{I}(\varphi ),
\label{sumio}
\end{equation}
where the $\tilde{X}$s are power series in $A$ and can contain derivatives
acting on the $\tilde{N}$s. The terms with $n=0,1$ do not contribute to the
sum and can be dropped. Indeed, write 
\[
\exp \left( -S_{N}(\varphi ,N)+S(\varphi )\right) =\int [\mathrm{d}L]\exp
\left( \int T(L)+\int \mathbf{\tau }_{vJ}\mathcal{N}^{v}(L)\mathcal{O}_{%
\mathrm{R}}^{J}(\varphi )-\int L_{I}\tilde{N}^{I}\right) . 
\]
It is easy to check that the exponent of the right-hand side vanishes for $%
\tilde{N}^{I}=0$. To see this we must focus on connected diagrams that do
not have external $L$-legs. Since all vertices have at least two $L$-legs,
all such diagrams are at least one-loop, so they vanish. This proves that $%
S_{N}(\varphi ,N)=S(\varphi )$ when $N^{I}=\mathcal{O}_{\mathrm{R}%
}^{I}(\varphi )$, therefore the term with $n=0$ can be dropped from the sum
of (\ref{sumio}). Similarly, the derivative with respect to $N$, calculated
at $\tilde{N}^{I}=0$, collects the set of connected diagrams with one
external $L$-leg, which must also contain at least one loop. Thus, the terms
with $n=1$ of (\ref{sumio}) also vanish.

We conclude that $S_{N}$ has a structure similar to the structure (\ref
{struttu}) of $\Omega $: 
\begin{equation}
S_{N}(\varphi ,N)=S(\varphi )+\frac{1}{2}\int \tilde{N}^{I}A_{IJ}\tilde{N}%
^{J}+\sum_{n\geqslant 2}(A_{I_{1}J_{1}}\tilde{N}^{J_{1}})\cdots
(A_{I_{n}J_{n}}\tilde{N}^{J_{n}})\hspace{0.01in}\tilde{X}_{I}^{I_{1}\cdots
I_{n}}\mathcal{O}_{\mathrm{R}}^{I}(\varphi ).  \label{struttucl}
\end{equation}
This is the general structure of the classical, bare and renormalized
actions in the proper approach.

Let us compare this action with the action (\ref{sl}), which is written
using the ``improper variables'' $\varphi ,L$. The terms of $S_{L}$ linear
in $L_{I}$ and the terms of $S_{N}$ linear in $A_{IJ}N^{J}$ are multiplied
by (minus) the renormalized composite fields $\mathcal{O}_{\mathrm{R}%
}^{I}(\varphi )$, therefore allow us to identify them. The improvement terms 
\[
T(L)=\frac{1}{2}\int L_{I}(A^{-1})^{IJ}L_{J},\qquad T_{N}(\tilde{N})\equiv 
\frac{1}{2}\int \tilde{N}^{I}A_{IJ}\tilde{N}^{J}, 
\]
correspond to each other. Similarly, the terms $\int \hat{\tau}_{vI}\mathcal{%
N}^{v}\mathcal{O}_{\mathrm{R}}^{I}$ correspond to the last sum in (\ref
{struttucl}). The constants $\tilde{X}$ are equal to the $\tau $s plus
perturbative corrections. Clearly, there are as many $\tilde{X}$s as $\tau $%
s, so we can invert the $\tilde{X}$-$\tau $ relations and consider the $%
\tilde{X}$s as independent parameters. Expanding the monomials quadratically
proportional to $\tilde{N}$ using the same basis $\mathcal{N}^{v}$ we used
for the monomials quadratically proportional to $L$, we conclude that the
most general proper classical action $S_{cN}$ has the form 
\begin{equation}
S_{cN}(\varphi ,N)=S_{c}(\varphi )+\frac{1}{2}\int \tilde{N}_{c}^{I}A_{IJ}%
\tilde{N}_{c}^{J}+\int \rho _{vI}\mathcal{N}^{v}(\tilde{N}_{c})\mathcal{O}%
_{c}^{I}(\varphi ),  \label{mogecl}
\end{equation}
where $\rho _{vI}$ are constants and $\tilde{N}_{c}^{I}=N^{I}-\mathcal{O}%
_{c}^{I}(\varphi )$. The proper renormalized action is then 
\begin{equation}
S_{N}(\varphi ,N)=S(\varphi )+\frac{1}{2}\int \tilde{N}^{I}A_{IJ}\tilde{N}%
^{J}+\int \hat{\rho}_{vI}\mathcal{N}^{v}(\tilde{N})\mathcal{O}_{\mathrm{R}%
}^{I}(\varphi ),  \label{moge}
\end{equation}
where $\hat{\rho}_{vI}=\rho _{vI}$ plus perturbative corrections. Recall
that all counterterms of type $T_{N}(\tilde{N})$ are moved to $\int \hat{\rho%
}_{v0}\mathcal{N}^{v}(\tilde{N})$, so the matrix $A$ is unrenormalized.

From (\ref{assigne1}), we find that the perturbative expansion is correctly
organized if we assume that the constants $\rho _{vI}$ are $\mathcal{O}%
(\delta ^{n_{I}-n_{v}-1})$, where $n_{v}$ is the $\delta $-degree of $%
\mathcal{N}^{v}(\tilde{N})$.

\subsection{Changes of variables in the proper action}

Now we study how the proper action $S_{N}$ transforms under a change of
variables. Inserting (\ref{baremu}) into (\ref{propz}) the identity $%
W(J,L)=W^{\prime }(J^{\prime },L^{\prime })$ follows defining 
\begin{equation}
\varphi ^{\prime }=\varphi +b_{I}N^{I},\qquad N^{I\hspace{0.01in}\prime
}=z_{J}^{I}N^{J},  \label{baremastra}
\end{equation}
which gives 
\[
S_{N}^{\prime }(\varphi ^{\prime },N^{\prime })=S_{N}(\varphi ,N),\qquad
\int J\varphi +\int L_{I}N^{I}=\int J^{\prime }\varphi ^{\prime }+\int
L_{I}^{\prime }N^{\prime \hspace{0.01in}I}. 
\]
As before, we have dropped the subscript $S$ in the integrated fields $%
N_{S}^{I}$.

We see that using the proper approach a change of variables (\ref{baremastra}%
) in the functional integral looks exactly as it looks in the master
functional, where we have formula (\ref{baremastr}). Enlarging the set of
integrated fields from $\varphi $ to the proper variables $\varphi ,N$ we
have linearized the change of variables also at the level of integrated
fields, and gained a lot of simplicity and clarity. We call (\ref{baremastra}%
) a \textit{proper field redefinition}.

Moreover, in the proper approach both the action $S_{N}(\varphi ,N)$ and the
term $\int J\varphi +\int L_{I}N^{I}$ behave as scalars, without talking to
each other. This means that a proper functional integral written in the
conventional form remains written that way at all stages of the variable
change. Because of this, replacements and true changes of variables are
practically the same thing. We recall that, instead, when we work with
improper variables, where we have only $\int J\varphi $ instead of $\int
J\varphi +\int L_{I}N^{I}$, lengthy procedures are necessary to retrieve the
conventional form after the change of variables \cite{fieldcov}.

Nevertheless, from (\ref{baremastra}) it is not evident what the $\varphi $%
-change of field variables truly is, once we eliminate the $N$s. To make it
more explicit it is sufficient to apply (\ref{baremastra}) and then
reconvert the transformed action into its proper form (\ref{moge}). The
operations necessary to achieve this goal are very similar to the
manipulations met in ref. \cite{fieldcov}, now viewed from the viewpoint of
the master functional.

Let $f(\varphi ^{\prime })=\varphi ^{\prime }+\mathcal{O}(b)$ denote the
recursive solution to the equation 
\begin{equation}
f(\varphi ^{\prime })=\varphi ^{\prime }-b_{I}\mathcal{O}_{\mathrm{R}%
}^{I}(f(\varphi ^{\prime })).  \label{recus}
\end{equation}
Using (\ref{baremastra}) and (\ref{recus}), we can write 
\begin{equation}
\varphi =f(\varphi ^{\prime })-b_{I}\bar{N}^{I},  \label{cus2}
\end{equation}
where 
\begin{equation}
\bar{N}^{I}\equiv N^{I}-\mathcal{O}_{\mathrm{R}}^{I}(f(\varphi ^{\prime })).
\label{nbs}
\end{equation}
We have 
\begin{equation}
\tilde{N}^{I}=N^{I}-\mathcal{O}_{\mathrm{R}}^{I}(\varphi )=N^{I}-\mathcal{O}%
_{\mathrm{R}}^{I}(f(\varphi ^{\prime })-b_{J}\bar{N}^{J})=\bar{N}^{I}+F^{I}(%
\bar{N},\varphi ^{\prime }),  \label{asw}
\end{equation}
where $F^{I}$ are local functions of order $b$ and order $\bar{N}$.

Inserting (\ref{asw}) and (\ref{cus2}) in $S_{N}(\varphi ,N)$ and expanding
in the basis of composite fields, we get 
\begin{equation}
S_{N}(\varphi ,N)=S(f(\varphi ^{\prime }))+\int \bar{N}^{I}E_{I}(f(\varphi
^{\prime }))+\frac{1}{2}\int \bar{N}^{I}\bar{A}_{IJ}\bar{N}^{J}+\int \bar{%
\rho}_{vI}\mathcal{N}^{v}(\bar{N})\mathcal{O}_{\mathrm{R}}^{I}(f(\varphi
^{\prime })),  \label{anbisso}
\end{equation}
where $\bar{A}_{IJ}=A_{IJ}+\mathcal{O}(b)$ and $\bar{\rho}_{vI}=\hat{\rho}%
_{vI}+\mathcal{O}(b)$ are new constants and $E_{I}$ are $\mathcal{O}(b)$%
-local composite fields proportional to (derivatives of) the field equations 
$\delta S(f(\varphi ^{\prime }))/\delta \varphi ^{\prime }$. For later
convenience, we focus our attention on $\delta S(f(\varphi ^{\prime
}))/\delta \varphi ^{\prime }$ rather than $\left. \delta S(\varphi )/\delta
\varphi \right| _{\varphi =f(\varphi ^{\prime })}$.

Formula (\ref{anbisso}) is not written in the form we want, since it
contains terms linear in $\bar{N}^{I}$. We must work out $\tilde{N}^{\prime 
\hspace{0.01in}I}=\bar{N}^{I}+\mathcal{O}(b)$, so that (\ref{anbisso}) turns
into the primed version of (\ref{moge}). A crucial fact is that the terms
linear in $\bar{N}^{I}$ are also proportional to the field equations of $%
S(f(\varphi ^{\prime }))$.

Calculate the derivative of (\ref{anbisso}) with respect to $\bar{N}$ and
set it to zero. This condition can be written as 
\[
\bar{N}^{I}=-(\bar{A}^{-1})^{IJ}E_{J}(f(\varphi ^{\prime }))-(\bar{A}%
^{-1})^{IJ}\bar{\rho}_{vK}\int \frac{\delta \mathcal{N}^{v}(\bar{N})}{\delta 
\bar{N}^{J}}\mathcal{O}_{\mathrm{R}}^{K}(f(\varphi ^{\prime })) 
\]
and solved recursively. The solution $\bar{N}^{I}=Y^{I}(\varphi ^{\prime })=%
\mathcal{O}(b)$ is local and proportional to the field equations $\delta
S(f(\varphi ^{\prime }))/\delta \varphi ^{\prime }$. Now, define 
\begin{equation}
\bar{N}^{\prime \hspace{0.01in}I}=\bar{N}^{I}-Y^{I}(\varphi ^{\prime })
\label{nby}
\end{equation}
and use this definition to replace $\bar{N}^{I}$ inside (\ref{anbisso}). We
get 
\[
S_{N}(\varphi ,N)=\bar{S}(\varphi ^{\prime })+\frac{1}{2}\int \bar{N}%
^{\prime \hspace{0.01in}I}\bar{A}_{IJ}^{\prime }\bar{N}^{\prime \hspace{%
0.01in}J}+\int \bar{\rho}_{vI}^{\prime }\mathcal{N}^{v}(\bar{N}^{\prime })%
\mathcal{O}_{\mathrm{R}}^{I}(f(\varphi ^{\prime })), 
\]
where $\bar{A}_{IJ}^{\prime }=A_{IJ}+\mathcal{O}(b)$ and $\bar{\rho}%
_{vI}^{\prime }=\rho _{vI}+\mathcal{O}(b)$ are new constants. The term
linear in $\bar{N}^{\prime }$ is absent by construction and 
\[
\bar{S}(\varphi ^{\prime })=\left. S_{N}(\varphi ,N)\right| _{\bar{N}%
=Y^{I}(\varphi ^{\prime })}=S(f(\varphi ^{\prime }))+\int \frac{\delta
S(f(\varphi ^{\prime }))}{\delta \varphi ^{\prime }}\mathcal{M}(f(\varphi
^{\prime }))\frac{\delta S(f(\varphi ^{\prime }))}{\delta \varphi ^{\prime }}%
, 
\]
where $\mathcal{M}(\varphi ^{\prime })=\mathcal{O}(b^{2})$ is local and can
contain derivatives acting to its left and to its right. Now we can apply
the theorem recalled in the appendix, which tells us that there exists a
perturbatively local function $g(\varphi ^{\prime })=\varphi ^{\prime }+%
\mathcal{O}(b^{2})$, such that 
\[
\bar{S}(\varphi ^{\prime })=S(f(g(\varphi ^{\prime }))). 
\]
Write 
\[
\varphi (\varphi ^{\prime })\equiv f(g(\varphi ^{\prime }))=\varphi ^{\prime
}-b_{I}\mathcal{O}_{\mathrm{R}}^{I}(\varphi ^{\prime })+\mathcal{O}(b^{2}). 
\]
This formula is the renormalized variable change associated with (\ref
{baremastra}). Inserting the inverse $\varphi ^{\prime }=\varphi ^{\prime
}(\varphi )$ of this relation in (\ref{nbs}) and (\ref{nby}), expanding in
the basis of composite fields, and then using the second of (\ref{baremastra}%
), we can write 
\begin{eqnarray*}
\bar{N}^{\prime \hspace{0.01in}I} &=&N^{I}-\mathcal{O}_{\mathrm{R}%
}^{I}(f(\varphi ^{\prime }))-Y^{I}(\varphi ^{\prime })=N^{I}-w_{J}^{I}%
\mathcal{O}_{\mathrm{R}}^{J}(\varphi (\varphi ^{\prime })) \\
&=&(z^{-1})_{J}^{I}\left( N^{\prime \hspace{0.01in}J}-(zw)_{K}^{J}\mathcal{O}%
_{\mathrm{R}}^{K}(\varphi (\varphi ^{\prime }))\right)
=(z^{-1})_{J}^{I}\left( N^{\prime \hspace{0.01in}J}-\mathcal{O}_{\mathrm{R}%
}^{\prime \hspace{0.01in}J}(\varphi ^{\prime })\right) =(z^{-1})_{J}^{I}%
\tilde{N}^{\prime \hspace{0.01in}J},
\end{eqnarray*}
where $w_{J}^{I}=\delta _{J}^{I}+\mathcal{O}(b)$ are constants and the
formula 
\begin{equation}
\mathcal{O}_{\mathrm{R}}^{\prime \hspace{0.01in}I}(\varphi ^{\prime
})=(zw)_{J}^{I}\mathcal{O}_{\mathrm{R}}^{J}(\varphi (\varphi ^{\prime }))
\label{bussik}
\end{equation}
tells us how the basis of composite fields is transformed by the change of
variables. Formula (\ref{bussik}) can also be used to work out how the
renormalization constants of composite fields are affected. Finally, 
\[
S_{N}(\varphi ,N)=S^{\prime }(\varphi ^{\prime })+\frac{1}{2}\int \tilde{N}%
^{\prime \hspace{0.01in}I}A_{IJ}^{\prime }\tilde{N}^{\prime \hspace{0.01in}%
J}+\int \rho _{vI}^{\prime }\mathcal{N}^{v}(\tilde{N}^{\prime })\mathcal{O}_{%
\mathrm{R}}^{\prime \hspace{0.01in}I}(\varphi ^{\prime })=S_{N}^{\prime
}(\varphi ^{\prime },N^{\prime }), 
\]
where $S^{\prime }(\varphi ^{\prime })=\bar{S}(\varphi ^{\prime })=S(\varphi
(\varphi ^{\prime }))$ is the transformed action and $A_{IJ}^{\prime
}=A_{IJ}+\mathcal{O}(b)$ and $\rho _{vI}^{\prime }=\rho _{vI}+\mathcal{O}(b)$
are new constants.

Observe that the procedure just described allows us to work out the
renormalization of the theory in the new variables without having to
calculate it anew. It is sufficient to know the renormalization (of the
action \textit{and} composite fields) in some variable frame to derive it in
any other variable frame using the change of variables.

We have learned that an operation as simple as (\ref{baremastra})
corresponds to a complex list of operations on the action. Nevertheless,
those operations are not completely new to us, since they resemble the
operations we had to do in ref. \cite{fieldcov} when we studied the changes
of field variables working with the $Z$- and $W$-functionals. These
observations show once again that the master functional is the correct
one-particle-irreducible partner of the $Z$- and $W$-functionals, while $%
\Gamma $ behaves in its own peculiar way.

\section{Renormalization of the master functional}

\setcounter{equation}{0}

In this section we study the renormalization of the master functional. We
first derive it from the renormalization of $W$. However, this method does
not make us appreciate the virtues of the master functional. Moreover, the
theorem of locality of counterterms can be applied in a much simpler way on
generating functionals of one-particle irreducible diagrams rather than on $%
W $. Therefore, we also derive the renormalization of $\Omega $ working
directly on $\Omega $, using the proper approach, without referring to the
definition of $\Omega $ from $W$.

The renormalization of $W$ in the linear redundant approach is encoded in
formula (7.14) of ref. \cite{fieldcov} and amounts to the source
transformation 
\begin{equation}
L_{I\mathrm{B}}=(L_{J}-\tilde{c}_{J}J)(\tilde{Z}^{-1})_{I}^{J},\qquad J_{%
\mathrm{B}}=J,  \label{BRrtasfa}
\end{equation}
plus parameter-redefinitions that we do not need to report here. Deriving
the renormalization of $\Omega $ from the one of $W$ is straightforward. The
transformation (\ref{BRrtasfa}) is a particular case of (\ref{baremu}), so
we know that it corresponds to a linear $\Phi $-$N$ redefinition of the form
(\ref{baremastr}) in $\Omega $ and an identical redefinition of the form (%
\ref{baremastra}) in the proper action $S_{N}(\varphi ,N)$.

This could be the end of the story, but we want to rederive these results
working directly on $\Omega $, to emphasize that the formulation of quantum
field theory using the master functional is completely autonomous. The
proper approach is very useful for our present purpose. If we forget about
the derivation just given, imported from the $W$-functional, it is not
obvious that the renormalization of $\Omega $ is just a linear redefinition
of the form (\ref{baremastra}) of the proper variables, plus a redefinition
of parameters. It is instructing to see how these properties emerge from $%
\Omega $.

As usual, we proceed inductively. We assume that renormalization works by
means of proper field redefinitions 
\[
\varphi \rightarrow \varphi +b_{I}N^{I},\qquad N^{I}\rightarrow
z_{J}^{I}N^{J}, 
\]
and parameter redefinitions up to $n$-loops and prove that then it works the
same way at $n+1$ loops. Call $\Omega _{n}$ the $\Omega $-functional
renormalized up to $n$ loops. Denote its proper fields with $\varphi _{n}$
and $N_{n}$, the parameters with $\lambda _{n}$ and $\rho _{n}$, the
composite fields with $\mathcal{O}_{n}^{I}(\varphi _{n})$ and the $n$-loop
renormalized proper action with $S_{N\hspace{0.01in}n}$. Using (\ref{moge}),
we can write 
\begin{equation}
S_{N\hspace{0.01in}n}(\varphi _{n},N_{n},\lambda _{n},\rho
_{n})=S_{n}(\varphi _{n},\lambda _{n},\rho _{n})+\frac{1}{2}\int \tilde{N}%
_{n}^{I}A_{IJ}\tilde{N}_{n}^{J}+\int \hat{\rho}_{vIn}\mathcal{N}^{v}(\tilde{N%
}_{n})\mathcal{O}_{\mathrm{R}n}^{I}(\varphi _{n}),  \label{garcia}
\end{equation}
where $\tilde{N}_{n}^{I}=N_{n}^{I}-\mathcal{O}_{\mathrm{R}n}^{I}(\varphi
_{n})$. As usual, we do not need to renormalize the constants $A_{IJ}$, as
counterterms for the improvement term are provided by $\int \hat{\rho}_{v0}%
\mathcal{N}^{v}(\tilde{N}_{n})$.

Recalling that the master functional is just the $\Gamma $-functional of the
proper variables, we can apply the theorem of locality of counterterms,
which tells us that the $(n+1)$-loop divergent part $\Omega _{n\hspace{0.01in%
}\text{div}}^{(n+1)}$ of $\Omega _{n}$ is a local functional. Organize $%
\Omega _{n\hspace{0.01in}\text{div}}^{(n+1)}$ as an expansion in powers of $%
\tilde{N}_{n}^{I}$: 
\begin{eqnarray*}
\Omega _{n\hspace{0.01in}\text{div}}^{(n+1)}(\varphi _{n},N_{n},\lambda
_{n},\rho _{n}) &=&\omega _{n}(\varphi _{n})+\int \frac{\delta S_{n}(\varphi
_{n})}{\delta \varphi _{n}}q_{In}\mathcal{O}_{\mathrm{R}n}^{I}(\varphi
_{n})+\int \tilde{N}_{n}^{I}\zeta _{IJn}\mathcal{O}_{\mathrm{R}%
n}^{J}(\varphi _{n}) \\
&&+\int \sigma _{vIn}\mathcal{N}^{v}(\tilde{N}_{n})\mathcal{O}_{\mathrm{R}%
n}^{I}(\varphi _{n}),
\end{eqnarray*}
where $q_{In}$, $\zeta _{IJn}$ and $\sigma _{vIn}$ are constants of order $%
(n+1)$-loop. We have separated the contributions at $\tilde{N}_{n}^{I}=0$
into two sets: the terms proportional to the field equations, whose
coefficients are also expanded in the basis $\mathcal{O}_{\mathrm{R}n}^{I}$
of composite fields, and the terms $\omega _{n}(\varphi )$ that must be
reabsorbed redefining the parameters $\lambda _{n}$ inside $S_{c}(\varphi )$%
. Now, the action $S_{N\hspace{0.01in}n+1}$ that renormalizes the theory up
to $n+1$ loops must be equal to $S_{N\hspace{0.01in}n}-\Omega _{n\hspace{%
0.01in}\text{div}}^{(n+1)}$ up to higher orders (which means $(n+2)$-loop or
higher), and its fields and parameters must then carry the subscript $n+1$.
We write 
\begin{equation}
S_{N\hspace{0.01in}n+1}(\varphi _{n+1},N_{n+1},\lambda _{n+1},\rho
_{n+1})=S_{N\hspace{0.01in}n}(\varphi _{n+1},N_{n+1},\lambda _{n+1},\rho
_{n+1})-\Omega _{n\hspace{0.01in}\text{div}}^{(n+1)}(\varphi
_{n+1},N_{n+1},\lambda _{n+1},\rho _{n+1}),  \label{jiuo}
\end{equation}
up to higher orders, which for the moment remain unspedified. It is clear
that the master functional $\Omega _{n+1}$ defined by the action (\ref{jiuo}%
) is convergent up to $n+1$ loops, since $\Omega _{n+1}=\Omega _{n}-\Omega
_{n\hspace{0.01in}\text{div}}^{(n+1)}$ up to that order. We want to show
that once field and parameters are converted to $\varphi _{n}$, $N_{n}$, $%
\lambda _{n}$ and $\rho _{n}$, by means of the proper field redefinitions 
\begin{equation}
\varphi _{n+1}=\varphi _{n}+q_{In}N_{n}^{I},\qquad
N_{n+1}^{I}=z_{nJ}^{I}N_{n}^{J},  \label{fnp0}
\end{equation}
and certain parameter redefinitions, 
\begin{equation}
\lambda _{n+1}=\lambda _{n}+\Delta _{n}\lambda _{n},\qquad \rho _{n+1}=\rho
_{n}+\Delta _{n}\rho _{n},  \label{fnpa}
\end{equation}
where the unknown constants $z_{nJ}^{I}-\delta _{J}^{I}$, $\Delta
_{n}\lambda _{n}$ and $\Delta _{n}\rho _{n}$ are $(n+1)$-loop, then the
right-hand side of formula (\ref{jiuo}) coincides with $S_{N\hspace{0.01in}%
n}(\varphi _{n},N_{n},\lambda _{n},\rho _{n})$ up to higher orders. Note
that we can also write 
\begin{equation}
\varphi _{n+1}=\varphi _{n}+q_{nI}\mathcal{O}_{\mathrm{R}n}^{I}(\varphi
_{n})+q_{nI}\tilde{N}_{n}^{I}.  \label{fnp1}
\end{equation}

The redefinitions of fields and parameters may be implemented writing 
\begin{eqnarray}
&&S_{N\hspace{0.01in}n}(\varphi _{n}+\Delta _{n}\varphi _{n},N_{n}+\Delta
_{n}N_{n},\lambda _{n}+\Delta _{n}\lambda _{n},\rho _{n}+\Delta _{n}\rho
_{n})=S_{N\hspace{0.01in}n}(\varphi _{n},N_{n},\lambda _{n},\rho _{n}) 
\nonumber \\
&&\qquad +\left( \int \Delta _{n}\varphi _{n}\frac{\delta }{\delta \varphi
_{n}}+\int \Delta _{n}N_{n}\frac{\delta }{\delta N_{n}}+\Delta _{n}\lambda
_{n}\frac{\partial }{\partial \lambda _{n}}+\Delta _{n}\rho _{n}\frac{%
\partial }{\partial \rho _{n}}\right) S_{cN}(\varphi _{n},N_{n},\lambda
_{n},\rho _{n})  \label{arci}
\end{eqnarray}
plus higher orders. In the corrections that appear on the right-hand side we
have replaced $S_{N\hspace{0.01in}n}$ with the classical proper action $%
S_{cN}$ (\ref{mogecl}). This is allowed since the difference is again made
of higher order terms.

As said, there must exist redefinitions $\lambda _{n+1}$ of the parameters $%
\lambda _{n}$ inside $S_{c}(\varphi )$ that reabsorb $\omega _{n}(\varphi
_{n})$. Then, using (\ref{arci}) and neglecting higher-orders, we can write
the right-hand side of (\ref{jiuo}) in the form 
\begin{eqnarray}
&&S_{N\hspace{0.01in}n}(\varphi _{n},N_{n},\lambda _{n},\zeta _{n})+\left(
\int \Delta _{n}\varphi _{n}\frac{\delta }{\delta \varphi _{n}}+\int \Delta
_{n}N_{n}\frac{\delta }{\delta N_{n}}+\Delta _{n}\rho _{n}\frac{\partial }{%
\partial \rho _{n}}\right) S_{cN}(\varphi _{n},N_{n},\lambda _{n},\rho _{n})
\nonumber \\
&&\qquad \qquad \qquad -\tilde{\Omega}_{n\hspace{0.01in}\text{div}%
}^{(n+1)}(\varphi _{n},N_{n},\lambda _{n},\rho _{n}).  \label{fnp2}
\end{eqnarray}
where 
\[
\tilde{\Omega}_{n\hspace{0.01in}\text{div}}^{(n+1)}(\varphi
_{n},N_{n},\lambda _{n},\rho _{n})=\int \frac{\delta S_{c}(\varphi _{n})}{%
\delta \varphi _{n}}q_{In}\mathcal{O}_{\mathrm{R}}^{I}(\varphi _{n})+\int 
\tilde{N}_{n}^{I}\tilde{\zeta}_{IJn}\mathcal{O}_{\mathrm{R}}^{J}(\varphi
_{n})+\int \tilde{\sigma}_{vIn}\mathcal{N}^{v}(\tilde{N}_{n})\mathcal{O}_{%
\mathrm{R}}^{I}(\varphi _{n}). 
\]
The constants in front of the last two divergent terms have been modified,
since the $\lambda _{n}$-redefinitions applied to (\ref{mogecl}) may also
affect those terms if the composite fields depend on $\lambda $. Thus, (\ref
{fnp2}) becomes 
\begin{equation}
S_{N\hspace{0.01in}n}(\varphi _{n},N_{n},\lambda _{n},\zeta _{n})-\int 
\tilde{N}_{n}^{I}\bar{\Delta}_{nIJ}\mathcal{O}_{\mathrm{R}}^{J}(\varphi
_{n})+\int (\Delta _{n}\rho _{vIn}-\bar{\sigma}_{vIn})\mathcal{N}^{v}(\tilde{%
N}_{n})\mathcal{O}_{\mathrm{R}}^{I}(\varphi _{n}),  \label{busto}
\end{equation}
plus higher orders, where 
\[
\bar{\Delta}_{nIJ}=A_{IK\hspace{0.01in}}(1-z_{n})_{J}^{K}+\rho _{vK\hspace{%
0.01in}n}C_{LIJ}^{vKM}(1-z_{n})_{M}^{L}+d_{IJ} 
\]
and $C_{LIJ}^{vKM}$, $d_{IJ}$ and $\bar{\sigma}_{vIn}$ are $(n+1)$-loop $%
\Delta _{n}\rho _{vIn}$-independent constants, $C_{LIJ}^{vKM}$ and $d_{IJ}$
being also $z_{n}$-independent. Finally, we can choose $z_{n}$ so that $\bar{%
\Delta}_{nIJ}=0$ and set $\Delta _{n}\rho _{vIn}=\bar{\sigma}_{vIn}$. Then (%
\ref{jiuo}) coincides with $S_{N\hspace{0.01in}n}(\varphi _{n},N_{n},\lambda
_{n},\zeta _{n})$ up to higher orders, which is the desired result.

Now we can upgrade formula (\ref{jiuo}), where higher-order contributions
remained unspecified, and define $S_{N\hspace{0.01in}n+1}$ by the exact
identity 
\begin{equation}
S_{N\hspace{0.01in}n+1}(\varphi _{n+1},N_{n+1},\lambda _{n+1},\zeta
_{n+1})=S_{N\hspace{0.01in}n}(\varphi _{n},N_{n},\lambda _{n},\zeta _{n}).
\label{putativo}
\end{equation}
This formula encodes the correct order-by-order renormalization, made of
proper field redefinitions (\ref{baremastra}) and parameter redefinitions.

We conclude that renormalization can be worked out directly on the master
functional following rules entirely similar to the ones we are accustomed
to. The advantage is that now we have a general field-covariant approach.
Moreover, all field redefinitions, including those that are part of the BR
map, are linear and there is no practical difference between replacements
and true changes of field variables.

\section{Generalizations}

\setcounter{equation}{0}

The master functional, as defined so far, is well suited for the linear
approach. There all changes of field variables, including the BR map, are
simple linear redefinitions of $\Phi $ and $N$. We have pointed out that the
Legendre transform is indeed invariant only under linear transformations.
Nevertheless, in ref. \cite{fieldcov} we have also been able to work with
the essential approach in the $W$-functional, and in section 2 we have been
able to do that in the $\Gamma $-functional. Thus, it must be possible to
generalize the definition of master functional to make it work with the most
general approach and the most general redefinitions of $\Phi $ and $N$. In
this section we elaborate a little bit on this issue.

Let us go back to formula (\ref{pegepo}). We have pointed out that its lack
of covariance is due to the fact that $x^{\mu }$ does not transform as a
vector under general coordinate transformations. Let us define a more
general transform, where $x^{\mu }$ is replaced by a vector $v^{\mu }(x)$.
We have 
\[
g(y)=-f(x)+v^{\mu }(x)\frac{\mathrm{d}f}{\mathrm{d}x^{\mu }},\qquad y^{\mu
}(x)=\frac{\mathrm{d}f}{\mathrm{d}x^{\mu }}. 
\]
Now $g(y)$ does transform correctly as a scalar, if $f$ does.

Let $V_{I}(J,L)$ denote perturbatively local functions of the sources. In
general, we assume that $V_{I}$ is equal to $L_{I}$ plus a perturbative
series in some expansion parameters. We call such parameters $\kappa $.
Moreover, we assume that $V_{I}$ is a vector in source space, which means
that it transforms as 
\begin{equation}
V_{I}^{\prime }=\int V_{J}\frac{\delta L_{I}^{\prime }}{\delta L_{J}}+\int J%
\frac{\delta L_{I}^{\prime }}{\delta J},  \label{forza}
\end{equation}
under a perturbatively local change of variables (\ref{bu}).

Define $\Phi $ and $N$ as in (\ref{relaf0}), but replace the definition (\ref
{definemast}) of the master functional with 
\begin{equation}
\Omega (\Phi ,N)=-W(J,L)+\int J\Phi +\int V_{I}N^{I}.  \label{definemast2}
\end{equation}
On $\Phi $ and $N$ the change of variables reads 
\[
\Phi ^{\prime }=\Phi +\int N^{I}\frac{\delta L_{I}}{\delta J^{\prime }}%
,\qquad N^{I\hspace{0.01in}\prime }=\int \frac{\delta L_{J}}{\delta
L_{I}^{\prime }}N^{J}, 
\]
where however $L^{\prime }$ and $J^{\prime }$ must still be replaced by the
appropriate functions of $\Phi $ and $N$. Since the relations $J(\Phi ,N)$
and $L(\Phi ,N)$ are in general non-local, the change of variables is
non-local in the space $\Phi $, $N$. Of course, it must be the sum of local
tree-level functions plus radiative corrections. We have 
\[
\Omega ^{\prime }(\Phi ^{\prime },N^{\prime })=\Omega (\Phi ,N), 
\]
as desired. We can also write 
\[
\Omega (\Phi ,N)=\Gamma (\Phi ,L)+\int V_{I}N^{I}. 
\]
Since $\Gamma (\Phi ,L)$ collects one-particle irreducible diagrams, and $%
V_{I}=L_{I}$ plus local perturbative corrections, $\Omega (\Phi ,N)$ also
collects one-particle irreducible diagrams. Nevertheless, in general $\Omega 
$ does not have the typical structure (\ref{gastru}), in the sense that its
radiative corrections do not follow from its classical limit with the usual
rules, and the classical limit of $\Omega $ is not necessarily the classical
action.

For example, we can take $V_{I}=L_{I}$ in the essential frame, which is the
variable frame where the action does not contain terms proportional to the
field equations, apart from those containing the free kinetic terms \cite
{fieldcov}. Then $\Omega $ is the Legendre transform of $W$ with respect to $%
J$ and $L$ in the essential frame, and has the structure (\ref{gastru}). In
every other frame we define $V_{I}$ as given by (\ref{forza}). With this
convention the vectors $V_{I}$ are inherited by a change of variables from
the essential frame.

The inverse formulas read 
\begin{equation}
J=\frac{\delta \Omega }{\delta \Phi }-\frac{\delta }{\delta \Phi }\int
(V_{I}-L_{I})N^{I},\qquad L_{I}=\frac{\delta \Omega }{\delta N^{I}}-\frac{%
\delta }{\delta N^{I}}\int (V_{J}-L_{J})N^{J}.  \label{lacce}
\end{equation}
If $\Omega $ were a Legendre transform its inverse would be a Legendre
transform. Instead, the procedure to obtain $W$ from $\Omega $ is more
complicated, and we cannot implement it unless we know the vector $%
V_{I}(J,L) $. Assuming that we have this knowledge, and recalling that $%
V_{I}-L_{I}=\mathcal{O}(\kappa )$, we can solve formulas (\ref{lacce})
recursively in powers of $\kappa $. This procedure gives us the functions $%
J(\Phi ,N)$ and $L_{I}(\Phi ,N)$. Once we have them we are ready to invert (%
\ref{definemast2}) and find 
\[
W(J,L)=-\Omega (\Phi ,N)+\int J\Phi +\int V_{I}N^{I}. 
\]

A similar procedure can be used to extract the expectation values of
elementary and composite fields from the master functional. These are the
constant solutions of the conditions $J(\Phi ,N)=L_{I}(\Phi ,N)=0$. Formulas
(\ref{lacce}) give 
\[
\frac{\delta \Omega }{\delta \Phi }=\frac{\delta }{\delta \Phi }\int
(V_{I}-L_{I})N^{I},\qquad \frac{\delta \Omega }{\delta N^{I}}=\frac{\delta }{%
\delta N^{I}}\int (V_{J}-L_{J})N^{J}. 
\]
Since the right-hand sides are $\mathcal{O}(\kappa )$, these equations can
be solved recursively in powers of $\kappa $. The zeroth-order expectation
values are the constant solutions of $\delta \Omega /\delta \Phi =\delta
\Omega /\delta N^{I}=0$.

\section{Conclusions}

\setcounter{equation}{0}

In this paper we have defined and studied a new generating functional of
one-particle irreducible diagrams, called master functional, which is
invariant with respect to the most general perturbative changes of field
variables.

A perturbative change of field variables starts with a redefinition of the
fields $\varphi $ in the action $S$. Inside the functionals $Z(J,L)$ and $%
W(J,L)$ it becomes a local perturbative redefinition of the sources $J$ and $%
L$ coupled to elementary and composite fields, under which $Z$ and $W$
behave as scalars. In a particularly convenient approach, the linear one,
such a source redefinition is linear. The functional $\Gamma (\Phi ,L)$, on
the other hand, does not behave as a scalar under the transformation law
inherited from its very definition. Nevertheless, there exists an unusual
field transformation under which $\Gamma $ does behave as a scalar. Instead,
the master functional $\Omega (\Phi ,N)$ behaves as a scalar under the
transformation law derived from its very definition, which is linear in $%
\Phi $ and $N$. We have worked out the relations among these three ways to
describe changes of field variables in quantum field theory and studied the
BR map as a particular case.

One obstruction to construct the master functional was that the Legendre
transform of $W$ with respect to the sources $L$ does not exist, in general.
We have solved this problem adding a certain ``improvement term'' to the
functional $W$, which equips the sources $L$ with suitable quadratic terms.
Then the master functional $\Omega (\Phi ,N)$ is defined as the Legendre
transform of the improved $W(J,L)$ with respect to both $J$ and $L$. We must
organize the perturbative expansion so that the ``$L$-propagators'' are
equal to unity. Then the master functional collects one-particle irreducible
diagrams. The lack of covariance of the Legendre transform is naturally
overcome in the linear approach, where all field redefinitions, including
those of the BR map, can be expressed linearly.

The master functional admits a very economic ``proper formulation'', where
the set of integrated fields is extended from $\varphi $ to the proper
variables $\varphi $-$N^{I}$, the $N^{I}$s being partners of the sources $%
L_{I}$ for composite fields. In this formulation the master functional is
the ordinary $\Gamma $-functional for the proper variables. The proper
classical action coincides with the classical limit of the master functional
and radiative corrections are the one-particle irreducible Feynman diagrams
of the proper formulation. Thus, they can be calculated working directly on
the master functional, without passing through $Z$, $W$ or $\Gamma $.
Finally, the conventional form of the functional integral is manifestly
preserved during a general change of field variables, so replacements and
true changes of field variables are practically the same thing.

An interesting subject for a future investigation is the generalization to
non-perturbative changes of field variables, which we have not considered
here.

\vskip 25truept \noindent {\Large \textbf{Appendix\quad Field redefinitions
and field equations}}

\vskip 15truept

\renewcommand{\theequation}{A.\arabic{equation}} \setcounter{equation}{0}

We know that if we perturb the action adding a local term proportional to
the field equations, we can reabsorb such a term inside the action by means
a local field redefinition to the first order of the Taylor expansion. It is
interesting to know that if we perturb the action adding a local term
quadratically proportional to the field equations, we can perturbatively
reabsorb it inside the action to \textit{all} orders by means of a local
field redefinition. In this appendix we briefly rederive this result and its
generalization to non-local functionals and non-local field redefinitions.
The theorem was proved in ref. \cite{acaus}, where a number of applications
and explicit examples can be found.

\begin{theorem}
Consider an action $S$ depending on fields $\phi _{i}$, where the index $i$
labels both the field type, the component and the spacetime point. Add a
term quadratically proportional to the field equations $S_{i}\equiv \delta
S/\delta \phi _{i}$ and define the modified action 
\begin{equation}
S^{\prime }(\phi _{i})=S(\phi _{i})+S_{i}F_{ij}S_{j},  \label{ayu}
\end{equation}
where $F_{ij}$ is symmetric and can contain derivatives acting to its left
and to its right. Summation over repeated indices (including the integration
over spacetime points) is understood. Then there exists a field redefinition 
\begin{equation}
\phi _{i}^{\prime }=\phi _{i}+\Delta _{ij}S_{j},  \label{redef}
\end{equation}
with $\Delta _{ij}$ symmetric, such that, perturbatively in $F$ and to all
orders in powers of $F$, 
\begin{equation}
S^{\prime }(\phi _{i})=S(\phi _{i}^{\prime }).  \label{equa}
\end{equation}
\end{theorem}

\textit{Proof}. The condition (\ref{equa}) can be written as 
\[
S(\phi _{i})+S_{i}F_{ij}S_{j}=S(\phi _{i}+\Delta _{ij}S_{j})=S(\phi
_{i})+\sum_{n=1}^{\infty }\frac{1}{n!}S_{k_{1}\cdots
k_{n}}\prod_{l=1}^{n}(\Delta _{k_{l}m_{l}}S_{m_{l}}), 
\]
after a Taylor expansion, where $S_{k_{1}\cdots k_{n}}\equiv \delta
^{n}S/(\delta \phi _{k_{1}}\cdots \delta \phi _{k_{n}})$. This equality is
verified if 
\begin{equation}
\Delta _{ij}=F_{ij}-\Delta _{ik_{1}}\left[ \sum_{n=2}^{\infty }\frac{1}{n!}%
S_{k_{1}k_{2}k_{3}\cdots k_{n}}\prod_{l=3}^{n}(\Delta
_{k_{l}m_{l}}S_{m_{l}})\right] \Delta _{k_{2}j},  \label{genfor}
\end{equation}
where the product is meant to be equal to unity when $n=2$. Equation (\ref
{genfor}) can be solved recursively for $\Delta $ in powers of $F$. The
first terms of the solution are 
\begin{equation}
\Delta _{ij}=F_{ij}-\frac{1}{2}F_{ik_{1}}S_{k_{1}k_{2}}F_{k_{2}j}+\cdots
\label{32}
\end{equation}

This result is very general. It works both for local and non-local theories.
If $S(\phi _{i})$ and $F_{ij}$ are perturbatively local, namely they can be
perturbatively expanded so that every order of the expansion is local, the
field redefinition (\ref{redef}) and the action $S^{\prime }(\phi _{i})$ are
perturbatively local. If both $S(\phi _{i})$ and $F_{ij}$ are local, in
general (\ref{redef}) and $S^{\prime }(\phi _{i})$ are only perturbatively
local. Actually, the resummation of the expansion can produce a non-local
field redefinition. Finally, if $S(\phi _{i})$ and $F_{ij}$ are local or
perturbatively local at the classical level, then (\ref{redef}) and $%
S^{\prime }(\phi _{i})$ are perturbatively local at the classical level.

\end{document}